\documentclass[preprint,12pt]{elsarticle}

\usepackage{amssymb}
\usepackage{lineno}
\usepackage{multirow}
\usepackage{float}
\usepackage{glossaries}
\usepackage[ruled,vlined,linesnumbered,resetcount]{algorithm2e}
\usepackage{xcolor}
\usepackage{hyperref}
\usepackage[group-separator={,}]{siunitx}
\usepackage{verbatim}

\journal{Elsevier}

\begin{document}

\begin{frontmatter}

\title{Inverse deformation analysis: an experimental and numerical assessment using the FEniCS Project \tnoteref{funding}
}
\tnotetext[funding]{This study was supported by European Union’s Horizon 2020 research and innovation programme under the Marie Sklodowska-Curie grant agreement No. 764644, No. 798244 and the financial support of the European Research Council Starting Independent Research Grant (ERC StG grant agreement No. 279578). Jack S. Hale is supported by the National Research Fund, Luxembourg, and cofunded under the Marie Curie Actions of the European Commission (FP7-COFUND) Grant No.~6693582.}

\author[add1]{Arnaud Mazier}
\ead{mazier.arnaud@gmail.com}
\author[add1]{Alexandre Bilger}
\ead{bilger.alexandre@gmail.com}
\author[add2,add3]{Antonio E. Forte}
\ead{aeforte@seas.harvard.edu}
\author[add4]{Igor Peterlik}
\ead{peterlik@gmail.com}
\author[add1]{Jack S. Hale}
\ead{jack.hale@uni.lu}
\author[add1,add5]{Stéphane P.A. Bordas\corref{cor}}
\ead{stephane.bordas@alum.northwestern.edu}

\cortext[cor]{Corresponding author}
\address[add1]{Institute of Computational Engineering, Department of Engineering, University of Luxembourg, 6, avenue de la Fonte, L-4364 Esch-sur-Alzette, Luxembourg.}
\address[add2]{Harvard University, 29 Oxford St, Cambridge MA 02138, USA.}
\address[add3]{Department of Electronics, Information and Bioengineering, Politecnico di Milano, Milan, 20133 Italy.}
\address[add4]{Institute of Computer Science, Masaryk University, Czech Republic.}
\address[add5]{Institute of Research and Development Duy Tan University, K7/25 Quang Trung, Danang, Vietnam.}

\begin{abstract}
In this paper we develop a framework for solving inverse deformation problems using the FEniCS Project finite element software. We validate our approach with experimental imaging data acquired from a soft silicone beam under gravity. In contrast with inverse iterative algorithms that require multiple solutions of a standard elasticity problem, the proposed method can compute the undeformed configuration by solving only one modified elasticity problem. This modified problem has complexity comparable to the standard one.
The framework is implemented within an open-source pipeline enabling the direct and inverse deformation simulation directly from imaging data. We use the high-level Unified Form Language (UFL) of the FEniCS Project to express the finite element model in variational form and to automatically derive the consistent Jacobian. Consequently, the design of the pipeline is flexible: for example, it allows the modification of the constitutive models by changing a single line of code. We include a complete working example showing the inverse deformation of a beam deformed by gravity as supplementary material.
\end{abstract}


\begin{highlights}
\item Inverse deformation algorithm can retrieve the undeformed configuration of soft objects 
\item Implementation of the inverse deformation algorithm in the FEniCS Project software 
\end{highlights}

\end{frontmatter}



\section{Introduction}
\textbf{Motivation.} The organization of a standard biomechanical \emph{deformation analysis} pipeline typically proceeds as follows. First, by using imaging techniques such as Magnetic Resonance Imaging (MRI) a segmented image of the region of interest is obtained. This segmented image is then meshed so that it can be used as input for a finite element simulation. The mesh is considered as the initial or undeformed (or reference) configuration of an elastic body. Then, by applying external forces to this elastic body we can find its deformed (or current) equilibrium configuration.

Conversely, an \emph{inverse deformation analysis} allows us to find the undeformed configuration of a body knowing its deformed configuration. In the case of an object subject to gravity, the undeformed configuration can be seen as a theoretical \emph{gravity-free} configuration. Consequently, determining the rest-position of an organ is of interest in many (bio)mechanical problems. For example, in abdominal aortic aneurysms to compute the residual stresses~\cite{Raghavan2006, Lu2007}, or in breast cancer as an intermedial configuration between the imaging and surgical stance~\cite{Mira2018}. Besides, this approach can also be used in problems of industrial interest such as tire or turbine blade design~\cite{Koishi2001,Cardona2008}. \smallbreak

\textbf{Problem statement.} The objective of inverse deformation analysis is to determine the undeformed configuration of an object such that it attains a known deformed configuration under the action of a known loading. It is important to note the distinction between inverse deformation analysis and common inverse problems. In a typical inverse problem, we might assume we know the applied forces, the initial and deformed configuration, and the goal is to determine the model parameters that minimize some distance (metric) between initial and deformed configurations. In an inverse deformation analysis, we assume we know the applied forces, boundary conditions, model parameters, and the deformed configuration. The objective is to determine the undeformed configuration that would lead to the deformed configuration if the external forces were to be applied.

\textbf{Background.} Several authors have tackled the problem of inverse deformation analysis using a variety of strategies. To the best of our knowledge~\cite{Zl1957} was the first to propose exchanging the role of the deformed and undeformed configurations, i.e. to express the displacement of the body as a function of the deformed state. The study was limited to plane strain deformations and uniform extension.~\cite{Schield1967} applied the same formalism to a homogeneous elastic material, without body force. He showed the equivalence of the equilibrium equations if the initial and deformed configurations are interchanged as well as the volumetric strain energies. The results provided by this approach are shown to be commensurate with those of~\cite{Zl1957} but are based on dual relations between the initial configuration and the deformed configuration.~\cite{Carlson1969} used a variational principle to achieve the same as~\cite{Schield1967} and showed the validity of the approach for different elastic materials. More recently~\cite{Carroll2005} mathematically analyzed the Schield transformation and the proven inverse deformation theorem. The theorem states that if a particular deformation is supported without body force for a specific strain energy $W$, then the inverse deformation is another energy $W^{*}$, derived from the first: $W^{*}(\boldsymbol{F})=\operatorname{det} (\boldsymbol{F}) W\left(\boldsymbol{F}^{-1}\right)$, where $\boldsymbol{F}$ is the deformation gradient.

~\cite{Govindjee1996,Govindjee1998} introduced the reparameterization of the weak form of the forward problem of finite elasticity as a solution method for the inverse problem. This approach only requires $C_{0}$ continuity and has a direct physical connection to the problem. Additionally, the procedure eliminates boundary condition difficulties, can be straightforwardly implemented using standard forward numerical methods, and can deal with both compressible or incompressible materials.

Inspired by~\cite{Govindjee1998} (Eulerian model) and~\cite{Yamada1998} (Arbitrary Lagrangian-Eulerian (ALE)),~\cite{Fachinotti2009} rewrote the constitutive equations in terms of Lagrangian variables. This manipulation makes the inverse analysis code changes limited to the finite element residual and Jacobian computations, contrary to Eulerian or ALE variables. The formulation is convenient and allows to solve \emph{inverse design problems} such as finding the unloaded shape of a turbine blade under known loading. But few drawbacks arise such as the difficulty of deriving and implementing the consistent Jacobian of the finite element formulation. Despite the usefulness of the approach, to our knowledge, this type of analysis is still not available in any widely used commercial simulation software.

Iterative methods identify the undeformed configuration based on several forward calculations. The algorithm is introduced by~\cite{Sellier2011} with a fixed-point method for elastostatic problems and then generalizes as the \emph{backward displacement method} by~\cite{Bols2013} for patient-specific blood vessel simulations. The iterative algorithm of Sellier has been widely applied to many image-based biomechanical simulations, mainly thanks to its algorithmic simplicity and its ability to use a standard non-linear elasticity simulation software~\cite{Mira2018}. However, when applied to strongly non-linear problems resulting from material or geometric non-linearity, the algorithm lacks robustness. Furthermore, iterative methods usually require at least one non-linear elasticity problem solution, resulting in higher costs compared with the approach of~\cite{Fachinotti2009}.

In the computer graphics community,~\cite{Chen2014} used Asymptotic Numerical Methods (ANM) to compute the rest-shape of elastic objects with a neo-Hookean material model. The ANM considers a parametrized version of the static equilibrium: $f(x, X)+\lambda g=0$, where $g$ is gravity, $\lambda$ a loading parameter and $f$ are the internal forces with the given deformed configuration $x$ and the unknown rest-configuration $X$. Then, the algorithm incrementally computes the asymptotic expansion of the curve in ($X, \lambda$) space until $\lambda = 1$, which corresponds to the rest-position. In this study, ANM offers superior performance, robustness, and convergence speed over traditional Newton-type methods for highly nonlinear material models. But the major drawback of the method is the complexity of changing the model formulation. Indeed, using a different material model implies to establish a different quadratic relationship between Cauchy stresses and the rest-position, then deriving the asymptotic local expansion. More recently,~\cite{Ly2018} developed an inversion algorithm applicable to geometrically non-linear thin shells, including the effects of contact and dry friction with an external body.

\textbf{Contribution.} In this paper we propose to use the Lagrangian formulation of~\cite{Fachinotti2009} coupled with automatic code generation tools provided by the FEniCS Project finite element software~\cite{alnaes_fenics_2015} to compute the rest or undeformed configuration of an object knowing the deformed configuration, the external loads and the material properties. We show experimental validation that the methodology is effective at recovering the undeformed configuration from imaging data. The formulation requires only a few minor modifications of the direct simulations, making it easy to implement. The automated differentiation tools from FEniCS Project provide a great deal of flexibility, for example, permitting users to quickly and easily modify the material model to suit their own problem. 

\textbf{Outline.} This paper is organized as follows; first, we give a description of the finite strain elasticity formulation and the constitutive equations used. Next, we explain the inverse deformation analysis method. We test our formulation on some simple analytical cases described in~\cite{Mihai2013,Lee2017}. Then, we show in some numerical examples how our variational formulation can surpass the iterative algorithm proposed by~\cite{Sellier2011}. Finally, we demonstrate a relevant real-world application by retrieving the undeformed configuration of a Polydimethylsiloxone (PDMS) beam under the action of gravity from imaging data.

\begin{figure}[H]
\includegraphics[width=\textwidth]{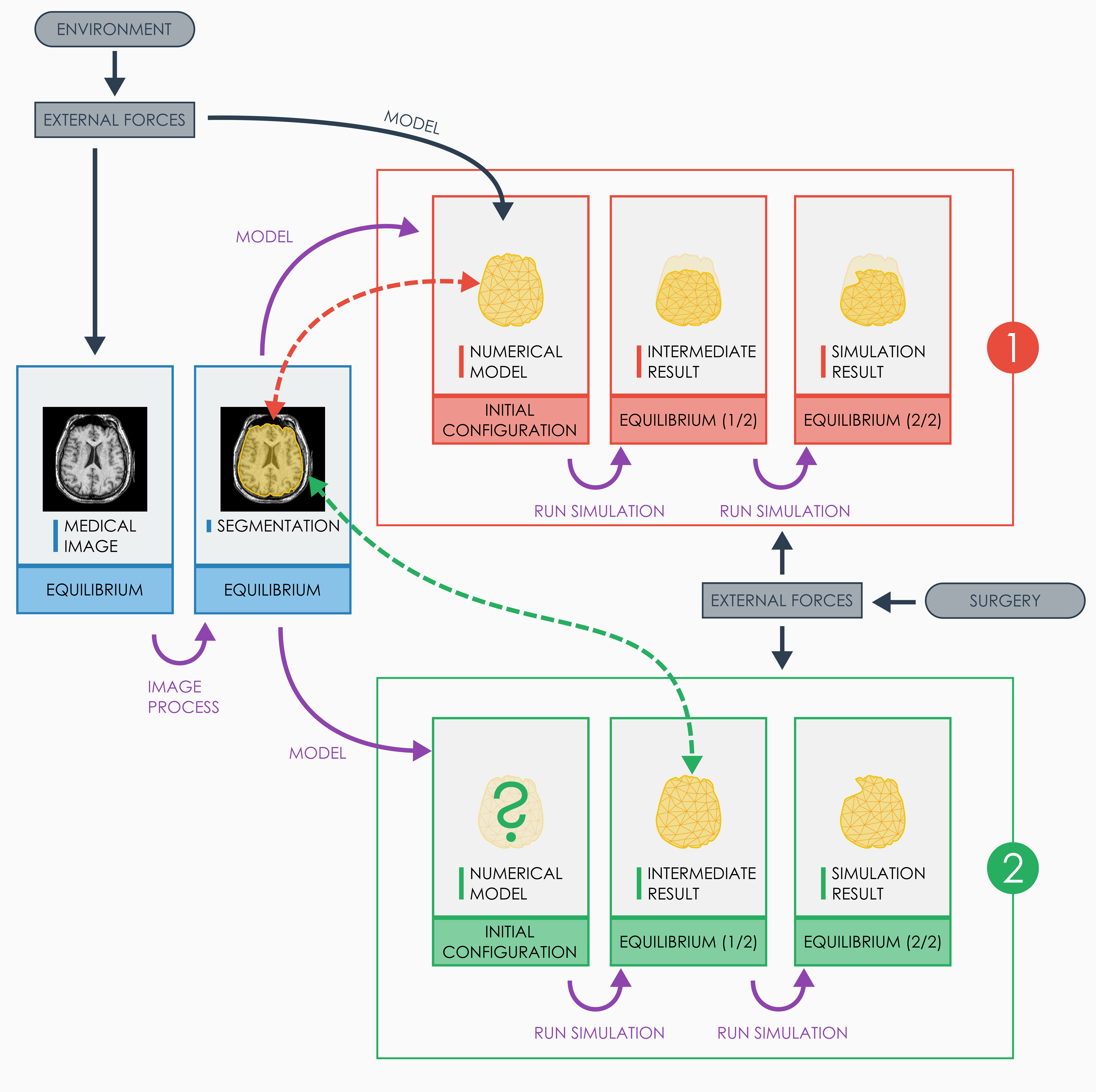}
\caption{Our pipeline starts with the acquisition of a medical image, in which the organ can be segmented. The organ is observed at equilibrium under the effect of external forces due to its environment (e.g. gravity). In the usual pipeline {\color{red}(1, in red)}, the segmented geometry is considered as the initial geometry. Then, the external forces are applied until equilibrium to obtain the intermediate geometry used for simulating the procedure. We propose an alternative approach {\color{green}(2, in green)}, where we constrain the intermediate geometry to be identical to the segmented geometry. It involves the computation of a new geometry (represented by the question mark symbol), which is the organ geometry such that it would deform to the segmented/imaged configuration if external forces were applied. Here, the final result takes into account the undeformed geometry of the organ.}
\label{fig:pipeline}
\end{figure}

\section{Finite strain elasticity formulation} \label{forward}
\subsection{Kinematics}

\begin{figure}[H]
\includegraphics[width=\textwidth]{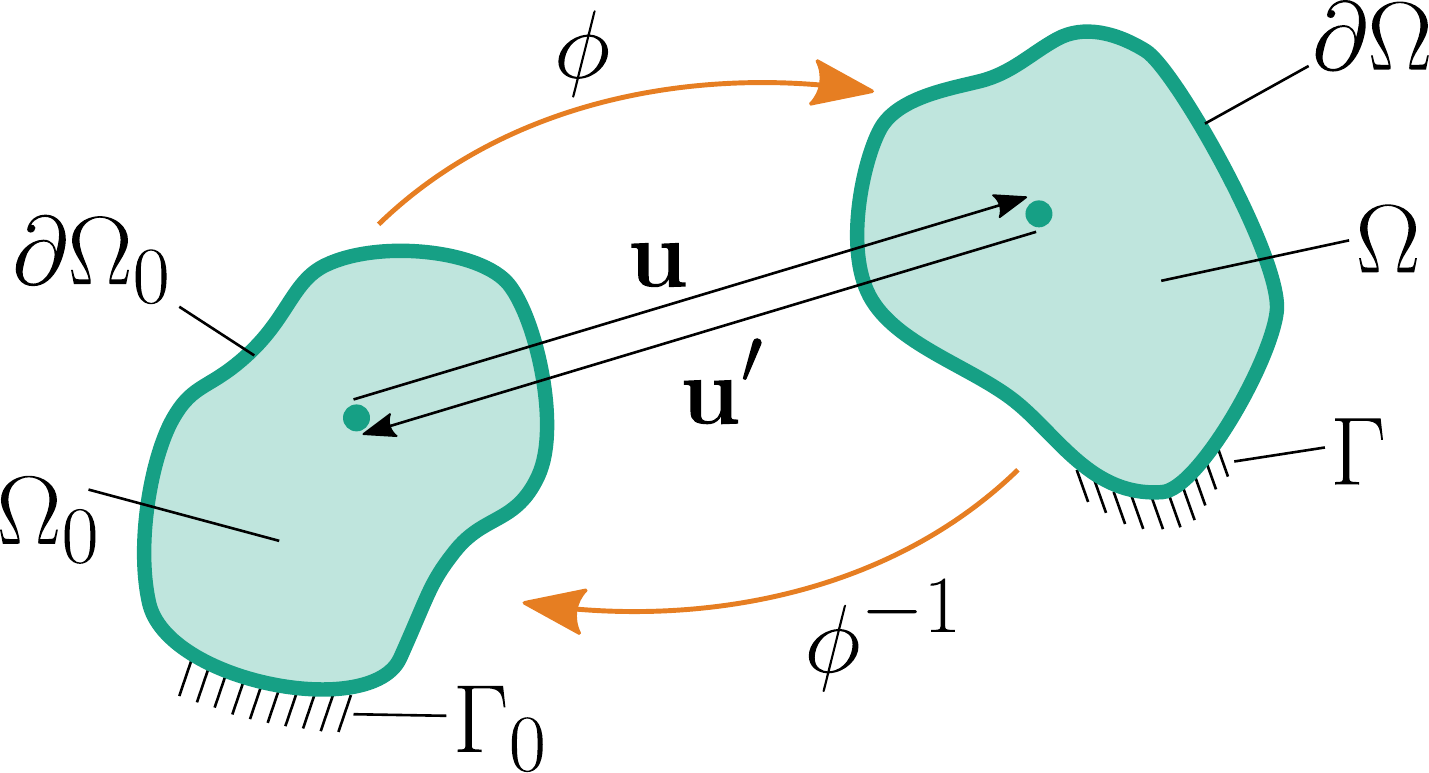}
\caption{In a standard deformation analysis we compute the displacement vector $\boldsymbol{u}$ from knowledge of the undeformed configuration $\Omega_0$. In the inverse deformation analysis the goal is to compute the displacement vector $\boldsymbol{u}^{'}$ from knowledge of the deformed configuration $\Omega$.}
\label{fig:directandinverse}
\end{figure}

Consider a deformable body $\mathcal{B}$. We denote the undeformed configuration $\Omega_{0}$. The location of a particle of $\mathcal{B}$ in $\Omega_{0}$ is denoted $\boldsymbol{X}$. Conversely, the deformed configuration is noted $\Omega$, and the location of a particle of $\mathcal{B}$ in $\Omega$ is noted $\mathrm{\mathbf{x}}$. A one-to-one mapping $\boldsymbol{\phi}$ maps the position of a particle $\boldsymbol{X}$ in $\Omega_{0}$ to the position of the same particle $\boldsymbol{x}$ in $\Omega$, i.e $\boldsymbol{x} = \phi(\boldsymbol{X})$.
The configuration $\Omega$ can be obtained by $\boldsymbol{\phi}\left(\Omega_{0}\right)=\left\{\boldsymbol{\phi}(\boldsymbol{X}) \; | \; \boldsymbol{X} \in \Omega_{0}\right\}$. These definitions are depicted in figure \ref{fig:directandinverse}. 

Let us introduce the deformation gradient $\boldsymbol{F}$ that maps a line element d$\boldsymbol{X}$ in $\Omega_{0}$ to a line element d$\boldsymbol{x}$ in $\Omega$:
	
\begin{equation} \label{eq:1}
d \boldsymbol{x}=\boldsymbol{F} \cdot d \mathbf{X}.
\end{equation}

We can write the deformation gradient and the Jacobian as
\begin{equation} \label{eq:2}
\boldsymbol{F}=\frac{\partial \boldsymbol{\phi}}{\partial \boldsymbol{X}},
\end{equation}
\begin{equation} \label{eq:3}
J=\operatorname{det} \boldsymbol{F}.
\end{equation}

As $\boldsymbol{\phi}$ is a one-to-one mapping, $\boldsymbol{F}$ is not singular and can be inverted, resulting in $J \ne 0$. The Jacobian maps a volume element $\mathrm{d}\Omega_{0}$ in $\Omega_{0}$ to a volume element $\mathrm{d}\Omega$ in $\Omega$
\begin{equation} \label{eq:4}
\mathrm{d}\Omega = J \cdot \mathrm{d}\Omega_{0}.
\end{equation}

For each point, we introduce the displacement $\boldsymbol{u}$ as the position difference between the deformed and the undeformed configuration
\begin{equation} \label{eq:5}
\begin{aligned} \boldsymbol{u} (\boldsymbol{X}) &=\boldsymbol{x}-\boldsymbol{X} \\
&=\boldsymbol{\phi}(\boldsymbol{X})-\boldsymbol{X} \end{aligned}
\end{equation}
The deformation gradient can also be written as a function of the displacement such as
\begin{equation} \label{eq:6}
\boldsymbol{F}=\frac{\partial \boldsymbol{\phi}}{\partial
	\boldsymbol{X}}=\frac{\partial \boldsymbol{u}}{\partial
	\boldsymbol{X}}+\boldsymbol{I}=\nabla_{0} \boldsymbol{u}+\boldsymbol{I},
\end{equation}
where $\nabla_{0}(\bullet)$ is the gradient in $\Omega_{0}$, with respect to the initial spatial position. The gradient in $\Omega$, with respect to the deformed spatial position, is denoted $\nabla(\bullet)$. $\boldsymbol{I}$ is the usual second-order identity tensor.
Similarly to the strain measure $\boldsymbol{F}$, we introduce the right Cauchy-Green strain tensor $\boldsymbol{C}$, its conjugate the left Cauchy-Green strain tensor $\boldsymbol{B}$ and the Green-Lagrange strain tensor $\boldsymbol{E}$
\begin{equation} \label{eq:7}
\mathbf{C}=\boldsymbol{F}^{T} \boldsymbol{F},
\end{equation}

\begin{equation} \label{eq:8}
\boldsymbol{B}=\boldsymbol{F} \boldsymbol{F}^{T},
\end{equation}

\begin{equation}  \label{eq:12}
\boldsymbol{E} =\frac{1}{2}(\mathbf{C}-\boldsymbol{I}).
\end{equation}

Hyperelastic material laws commonly use invariants of $\boldsymbol{C}$ and $\boldsymbol{B}$ to define their elastic energy
\begin{align} 
\mathrm{I}_{\mathrm{\boldsymbol{C}}} &=\operatorname{tr}(\mathbf{C}) \label{eq:9}, \\
\mathrm{II}_{\mathrm{\boldsymbol{C}}}
&=\frac{1}{2}\left((\operatorname{tr}(\mathbf{C}))^{2}-\operatorname{tr}\left(\mathbf{C}^{2}\right)\right)
\label{eq:10}, \\
\mathrm{III}_{\mathrm{\boldsymbol{C}}} &=\operatorname{det} \mathbf{C} \label{eq:11}.
\end{align}
\subsection{Strong form}
At equilibrium in the deformed configuration, the balance of momentum can be written as follows
\begin{equation} \label{eq:14}
\nabla \cdot \boldsymbol{\sigma}+\rho \boldsymbol{b} = \boldsymbol{0},
\end{equation}
where $\boldsymbol{\sigma}$ is the Cauchy stress tensor, $\rho$ is the density of the material in the deformed configuration and $\boldsymbol{b}$ are the external forces in the deformed configuration. Equation \ref{eq:14} is called the strong form and is written in the deformed configuration $\Omega$. To write the strong form in the initial configuration $\Omega_{0}$, we introduce the density of the material $\rho_{0}$ in the undeformed configuration and the first Piola-Kirchhoff stress tensor $\boldsymbol{P}$
\begin{equation} \label{eq:15}
\nabla_{0} \cdot \boldsymbol{P}+\rho_{0} \boldsymbol{b} = \boldsymbol{0},
\end{equation}
where $\boldsymbol{\sigma}$ and $\boldsymbol{P}$ are related by the Piola transform
\begin{equation} \label{eq:16}
\boldsymbol{\sigma}=\frac{1}{J} \boldsymbol{P} \boldsymbol{F}^{T}.
\end{equation}

\subsection{Weak form}
The weak form is obtained by multiplying the strong form by test functions
$\boldsymbol{\eta}$ and integrating over the whole domain. Equation \ref{eq:15} is written in the initial configuration and leads to
\begin{equation} \label{eq:17}
- \int_{\Omega_{0}} (\nabla_{0} \cdot \boldsymbol{P}) \cdot \boldsymbol{\eta} \; \mathrm{d}
\Omega_{0} = \int_{\Omega_{0}} (\rho_{0} \boldsymbol{b}) \cdot \boldsymbol{\eta} \; \mathrm{d} \Omega_{0}.
\end{equation}

By using the divergence theorem we obtain
\begin{equation}
- \int_{\Omega_{0}} (\nabla_{0} \cdot \boldsymbol{P}) \cdot \boldsymbol{\eta} \; \mathrm{d}
\Omega_{0} = \int_{\Omega_{0}} \boldsymbol{P} : \nabla_{0} \boldsymbol{\eta} \; \mathrm{d}
\Omega_{0} - \int_{\partial\Omega_{0}} (\boldsymbol{P} \cdot \boldsymbol{n}) \cdot \boldsymbol{\eta} \; \mathrm{d} \partial\Omega_{0},
\end{equation}
where the colon operator $:$ is the inner product between tensors, $\boldsymbol{n}$ is the outward unit normal at the boundary and $\partial\Omega_{0}$ the surface boundary of $\Omega_{0}$. The quantity $\boldsymbol{P} \cdot \boldsymbol{n}$ is the traction boundary condition. We here assume that it is prescribed on a part $\Gamma_{0}$ of the boundary as $\boldsymbol{P} \cdot \boldsymbol{n} = \boldsymbol{t}_0$. On the remaining part of the boundary, we assume that the value of the displacement is given, i.e. a Dirichlet condition. We then obtain the equilibrium in the reference configuration
\begin{equation} \label{eq:18}
\int_{\Omega_{0}} \boldsymbol{P} : \nabla_{0} \boldsymbol{\eta} \; \mathrm{d}
\Omega_{0}
=\int_{\Omega_{0}} \rho_{0} \boldsymbol{b}_0 \cdot \boldsymbol{\eta} \; \mathrm{d}
\Omega_{0}+\int_{\Gamma_{0}} \boldsymbol{t}_0 \cdot \boldsymbol{\eta} \; \mathrm{d} \Gamma_{0}.
\end{equation}

Note that the boundary integral on the remaining part $\partial\Omega_{0} \setminus \Gamma_{0}$ vanishes due to the Dirichlet condition. By injecting equation \ref{eq:16} in the last equation \ref{eq:18}, we obtain the weak form in the deformed configuration
\begin{equation} \label{eq:19}
\int_{\Omega} \boldsymbol{\sigma} : \nabla \boldsymbol{\eta} \; \mathrm{d} \Omega
=\int_{\Omega} \rho \boldsymbol{b} \cdot \boldsymbol{\eta} \; \mathrm{d}
\Omega+\int_{\Gamma} \boldsymbol{t} \cdot \boldsymbol{\eta} \; \mathrm{d} \Gamma.
\end{equation}

\subsection{Constitutive models}
For many materials, simple elastic models such as the St.\ Venant Kirchhoff model are not sufficient to describe the observed behavior. More complex hyperelastic models provide a mechanism of modeling the stress-strain behavior of complex materials such as elastomers or biological tissues.
\subsubsection{Compressible models}
\medbreak
\textbf{Neo-Hookean}
A neo-Hookean solid is a hyperelastic material model that can be used for predicting the nonlinear stress-strain behavior of materials undergoing large deformations. Its strain energy density is defined as:
\begin{equation} \label{eq:22}
\psi_{\mathrm{NH}}=\frac{\mu}{2} (\mathrm{I}_{B}-3) - \mu \ln(J) + \frac{\lambda}{2} \ln(J)^{2},
\end{equation}
where $\lambda$ and $\mu$ are material constants  called the Lamé parameters.

\medbreak
\noindent\textbf{Mooney-Rivlin}
A Mooney–Rivlin solid is a hyperelastic material model where the strain energy density function $\psi_{\mathrm{MR}}$ is a linear combination of two modified invariants of the left Cauchy–Green deformation tensor $\boldsymbol{B}$. Rubber-like materials are often modeled using the Mooney–Rivlin model with strain energy density
\begin{equation} \label{eq:23}
\psi_{\mathrm{MR}}=C_{1}\left(\overline{\mathrm{I}_{B}}-3\right)+C_{2}\left(\overline{\mathrm{I
	I}_{B}}-3\right)+D_{1}(J-1)^{2},
\end{equation}
with the modified invariants $\overline{\mathrm{I}_{B}} = J^{-\frac{2}{3}} \; \mathrm{I}_{B}$, $\overline{\mathrm{II}_{B}} = J^{-\frac{4}{3}} \; \mathrm{II}_{B}$ and where $C_{1}$, $C_{2}$, $D_{1}$ are material constants. 

\subsubsection{Nearly-incompressible model variants} \label{fenics-model}

All material models previously introduced were intended for compressible materials, i.e. materials where the volume may change during deformation. Conversely, some materials such as living tissues or rubbers can be assumed to be nearly-incompressible or even completely incompressible, i.e. volume is preserved during deformation $J \sim 1$.

For a hyperelastic material, the strain energy density function describes the stored energy as a function of the isochoric deformation, i.e.\ shape deformations without volume change. But using the standard displacement-based finite element method to describe incompressible material behavior may cause numerical problems typically referred to as \emph{locking}. Simply put, locking occurs when too many constraints are imposed on the discrete formulation and its overall approximation power is destroyed.

To overcome these difficulties, mixed formulations have been developed. In these formulations, the variational principle is modified by writing the potential energy functional similar to equation \ref{eq:23}, except that the strain energy is expressed in terms of the deviatoric component only and the incompressibility constraint is explicitly enforced using a Lagrange multiplier with physical meaning akin to pressure ($p$). It turns out that the Lagrange multipliers can be expressed as a function of the hydrostatic pressure values $f(p)$. It can be shown that
\begin{equation}
\psi(\boldsymbol{u}, p)= \psi(\boldsymbol{u}) - f(p),
\end{equation}
\begin{equation}
    f(p):= -\sigma_{\text{hydro}} = -\frac{1}{3} \operatorname{tr}(\boldsymbol{\sigma})=-\frac{\partial \psi}{\partial J}.
\end{equation}
\noindent\textbf{Neo-Hookean} By calculating $f(p)$, we can deduce the mixed displacement-pressure formulation of a nearly-incompressible Neo-Hookean material
\begin{equation}
\psi_{\mathrm{NH}} (\boldsymbol{u},p) =\frac{\mu}{2} (\mathrm{I}_{B}-3) - \mu \ln(J) + p \ln(J) - \frac{1}{2\lambda}p^{2}.
\end{equation}
\noindent\textbf{Mooney-Rivlin} By calculating $f(p)$, we can deduce the mixed displacement-pressure formulation of a nearly-incompressible Mooney-Rivlin material
\begin{equation}
\psi_{\mathrm{MR}} (u,p)=C_{1}\left(\overline{\mathrm{I}_{B}}-3\right)+C_{2}\left(\overline{\mathrm{I}
	\mathrm{I}_{B}}-3\right)+ p(J-1) - \frac{1}{4D_{1}}p^{2}.
\end{equation}
These nearly-incompressible energy densities are used to generate the FEniCS Project results in this paper.

\subsection{Finite element solver}
We use the FEniCS Project finite element software~\cite{alnaes_fenics_2015} to discretise both the standard finite strain elasticity problem and the inverse finite strain elasticity problem that we will outline in the next section. We use a mixed displacement-pressure finite element formulation with second-order continuous Lagrangian finite elements for displacement $\mathbf{u}$ and first-order continuous Lagrangian finite elements for pressure $p$. This pairing is well-known to be $\inf$-$\sup$ stable and relatively robust with respect to numerical locking.

The variational forms of the residual equations \ref{eq:18} and \ref{eq:27} are defined in the Unified Form Language (UFL)~\cite{alnaes_unified_2014} and symbolically differentiated to derive an expression for consistent Jacobian. The FEniCS Form Compiler (FFC)~\cite{logg_ffc_2012} is used to automatically generate low-level C\texttt{++}  code from the high-level UFL description that can calculate the Jacobian and residual cell tensors. The overall solution process is driven by the DOLFIN finite element library~\cite{logg_dolfin:_2010}. We use a standard Newton-Raphson algorithm with continuation in the loading parameter. The linear system within the Newton-Raphson algorithm is solved using the direct solver MUMPS via PETSc~\cite{petsc-web-page}. The complete implementation of the standard or inverse problem is around 100 lines of Python code that closely follows the mathematical structure of the problem. We refer the reader to the supplementary material~\cite{my_code-web-page} for further details.

\section{Inverse finite strain elasticity formulation}
This section presents two methods to compute the undeformed configuration knowing the deformed configuration under known loading. We first introduce our methodology derived from~\cite{Fachinotti2009}, then we briefly outline a simple iterative geometric algorithm described in~\cite{Sellier2011}.

\subsection{Inverse method}
In section \ref{forward}, we introduced how to compute the deformed configuration of a body undergoing external forces. The inputs were the undeformed geometry and the external forces, which means $\boldsymbol{X}$, the rest-position was known and $\boldsymbol{x}$, the deformed position, was unknown. In this section, we introduce our method to compute the undeformed configuration of a body undergoing external forces. The inputs of the inverse deformation formulation are the deformed geometry and the forces applied to the body. The most intuitive approach is to solve equation \ref{eq:18} or equation \ref{eq:19} for the unknown $\boldsymbol{X}$. This approach has the advantage of being based on classical mechanical principles. However, mechanical quantities such as strains or stresses are defined depending on $\boldsymbol{X}$. This approach requires few straightforward modifications to the equations in order to solve equation \ref{eq:18} or \ref{eq:19}. In the inverse approach, the initial geometry is replaced by the deformed geometry ($\boldsymbol{x}$). We redefine the displacement of equation \ref{eq:5} as
\begin{equation}\label{eq:25}
\boldsymbol{u}^{\prime}(\boldsymbol{x})=\boldsymbol{X}-\boldsymbol{x}.
\end{equation}
Note that trivially
\begin{equation}
\boldsymbol{u}^{\prime} + \boldsymbol{u} = \boldsymbol{0}.
\end{equation}

This redefinition does not modify the classical finite element pipeline: the unknown position is still the first term in which the known position is subtracted. Notice the selection of the gradient compared to equation \ref{eq:6}: we now compute gradients in the deformed configuration and this necessitates the redefinition of the deformation gradient
\begin{equation}\label{eq:26}
\boldsymbol{F}=\frac{\partial \boldsymbol{\phi(\boldsymbol{X})}}{\partial \boldsymbol{X}}
=\left(\frac{\partial \boldsymbol{X}}{\partial \boldsymbol{x}}\right)^{-1}
=\left(\frac{\partial \boldsymbol{u'(x)}}{\partial \boldsymbol{x}} + \boldsymbol{I}\right)^{-1}
=\left(\nabla \boldsymbol{u}^{\prime}+\boldsymbol{I}\right)^{-1}.
\end{equation}
Henceforth, when performing an inverse deformation analysis, $\boldsymbol{F}$ and all derived quantities (strain measures, invariants, energy densities, stress measures etc.) are always computed using the above redefinition in terms of $\boldsymbol{u}^{\prime}$.

The goal then is to solve equation \ref{eq:18} with $\boldsymbol{x}$ known and $\boldsymbol{X}$ unknown for $\boldsymbol{u}^{\prime}$. The weak equilibrium in the inverse deformation is expressed in the deformed configuration
\begin{equation} \label{eq:27}
\int_{\Omega} \boldsymbol{\sigma} : \nabla \boldsymbol{\eta} \; \mathrm{d} \Omega=\int_{\Omega} \rho
\boldsymbol{b} \cdot \boldsymbol{\eta} d \Omega+\int_{\Gamma} \boldsymbol{t} \cdot \boldsymbol{\eta} \; \mathrm{d} \Gamma.
\end{equation}

We can notice three main differences compared to equation \ref{eq:17}: (1) The integration domain is no longer the undeformed domain but the deformed domain. (2) The gradient of the trial function is in the deformed configuration. (3) The external forces are written in the deformed configuration. This new formulation requires us to make one change compared to the direct pipeline; rewrite $\boldsymbol{F}$ in terms of $\boldsymbol{u}^{\prime}$.
The computation of the gradient and the integration domain in the deformed configuration in the inverse analysis is equivalent to the computation of the gradient and the integration domain in the undeformed configuration in the direct analysis. That is why these changes in the formulation do not require a significant modification of a code to perform the inverse analysis. This formulation can find the undeformed configuration of an object, knowing only the deformed configuration and the applied forces. The process is "one-shot" based on the equation of continuum mechanics.

\subsection{Iterative geometric algorithm} \label{iga}
~\cite{Sellier2011} proposed an Iterative Geometric Algorithm (IGA, not to be confused with Isogeometric Analysis). The algorithm is simple to implement and only requires an existing (standard) forward deformation solver.
The algorithm starts with an initial guess for the undeformed configuration (usually chosen, for lack of a better choice, the deformed one) and applies successive displacement fields to it until a convergence criterion is reached. The sequence of displacement fields is obtained from the direct simulations of the current rest-configuration undergoing external forces. The shape of the object after the direct simulation provides an error compared to the exact rest-configuration by measuring the distance to the initial configuration. An updated estimate of the undeformed configuration is calculated by correcting the previous guess with the difference between the computed and deformed configuration. The algorithm stops when the error (computed using the $l^{2}$-norm) is below a defined threshold $\epsilon$ or a maximum number of iterations $NB^{\mathrm{max}}$ has been reached. The process is outlined in algorithm \ref{algo2}.

\begin{algorithm}[H]
	\SetAlgoLined
	$X^{0} \leftarrow X^{\mathrm{ini}} $\\
	run direct simulation 0 with $X^{0}$ the initial configuration\\
	$u^{0}\leftarrow x^{0} - X^{0}$\\
	err $\leftarrow$ error between $x^{0}$ and $X^{\mathrm{ini}}$\\
	$j \leftarrow 1 $\\
	\While{$\mathrm{err} > \epsilon$ \textbf{and} $j<NB^{\mathrm{max}}$}{
		$X^{j}\leftarrow X^{(j-1)} - u^{(j-1)}$\\
		run direct simulation j with $X^{(j-1)}$ the initial configuration\\
		$u^{j}\leftarrow x^{j} - X^{j}$\\
		$\mathrm{err} \leftarrow$ error between $x^{j}$ and $X^{\mathrm{ini}}$\\
		$j \leftarrow j+1 $
	}
	\caption{Iterative geometric algorithm from~\cite{Sellier2011}.}
	\label{algo2}
\end{algorithm}

\section{Numerical results}\label{results}
\subsection{Verification of the direct simulation}
The inverse deformation framework is very similar to the traditional direct framework. To assess the numerical precision of the inverse method, we first apply a series of tests to verify the soundness of the direct approach in which an analytic solution is known.

\subsubsection{Shear deformation} \label{simple_shear}
\begin{figure}[H]
\includegraphics[width=\textwidth]{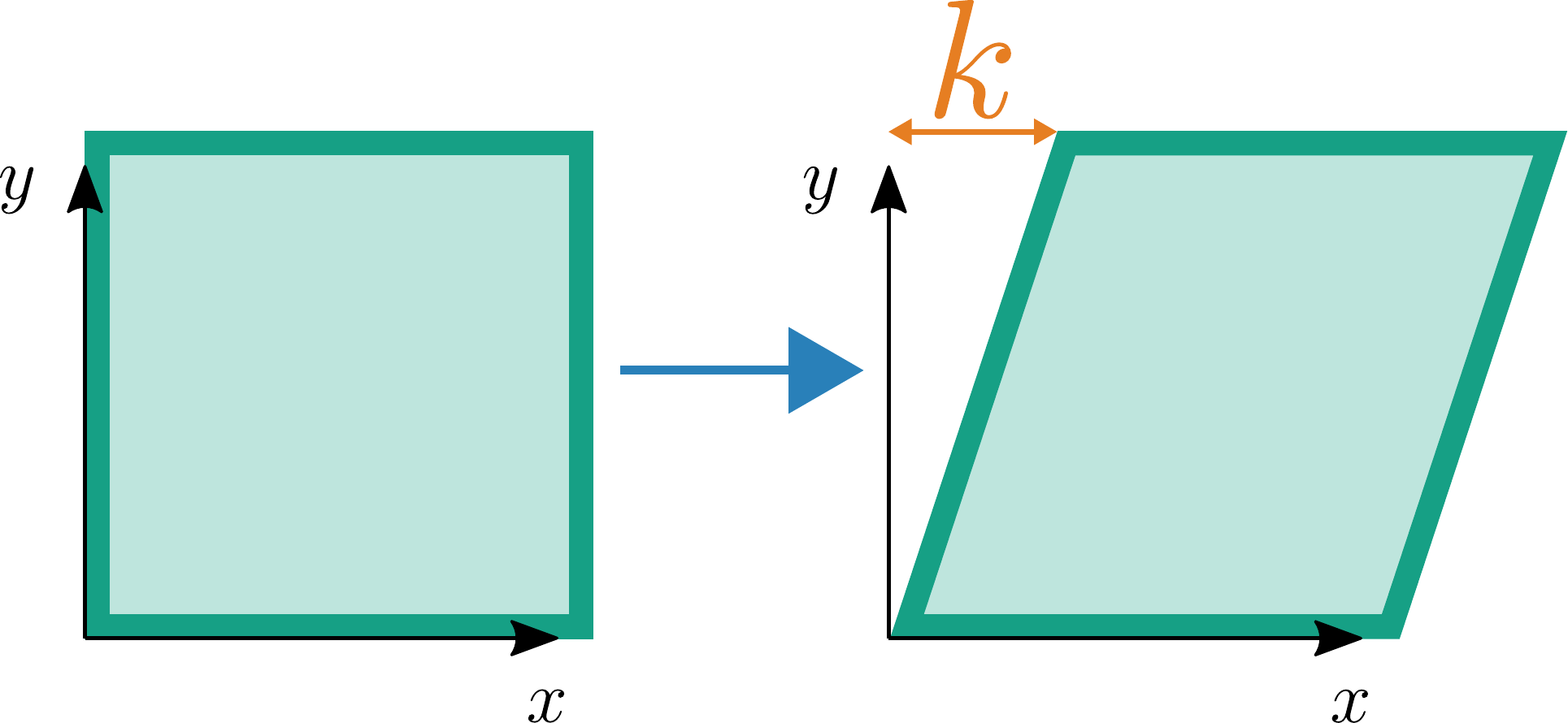}
\caption{2D plane cut of a simple shear deformation of a unit cube. An $x$-displacement of $y \cdot k$ is applied on the boundary.}
\label{fig:simple_shear}
\end{figure}

\textbf{Simple shear}: Simple shear deformation is a popular benchmark test~\cite{Mihai2013}. The initial geometry is a unit cube with prescribed Dirichlet boundary conditions $\boldsymbol{u}_0 = (y \cdot k, 0, 0)^T$ with $y$ the $y$-coordinate and $k$ a constant, as illustrated in figure \ref{fig:simple_shear}.

For simple shear deformation, the deformation gradient is equal to
\begin{equation} \label{eq:28}
\boldsymbol{F} =  \nabla_{0} \boldsymbol{u} + \boldsymbol{I} = 
\begin{pmatrix}
1 & k & 0 \\
0 & 1 & 0 \\
0 & 0 & 1
\end{pmatrix}.
\end{equation}
Now, let us consider a cube made of a Mooney-Rivlin material. By replacing the deformation gradient $\boldsymbol{F}$ in the equation \ref{eq:23}, we obtain the value of the strain energy density function in the cube. Following~\cite{Mihai2013} we can obtain the energy density and the components of the Cauchy stress tensor $\boldsymbol{\sigma}$
\begin{equation} \label{eq:29}
\psi = k^{2}(C_{1}+C_{2}),
\end{equation}
\begin{align} \label{eq:30}
\begin{split}
&\sigma_{00} = \frac{k^{2}(2C_{2}+4C_{1})}{3}, \\
&\sigma_{11} = -\frac{k^{2}(4C_{2}+2C_{1})}{3}, \\
&\sigma_{22} = \frac{k^{2}(2C_{2}-2C_{1})}{3}, \\
&\sigma_{01} = k(2C_{2}+2C_{1}), \\
&\sigma_{02} = \sigma_{12} = 0. \\
\end{split}
\end{align}
The values of $\psi$ and $\boldsymbol{\sigma}$ have been evaluated in our framework with several values of $k$, degrees of discretization, and constitutive parameters. The relative error (by using the $L^{2}$-norm) in strain energy and Cauchy stress tensor, compared to the analytical values, shows the exactness of the direct deformation framework to machine precision ($10^{-12}$ magnitude error).

\begin{figure}[H]
\includegraphics[width=\textwidth]{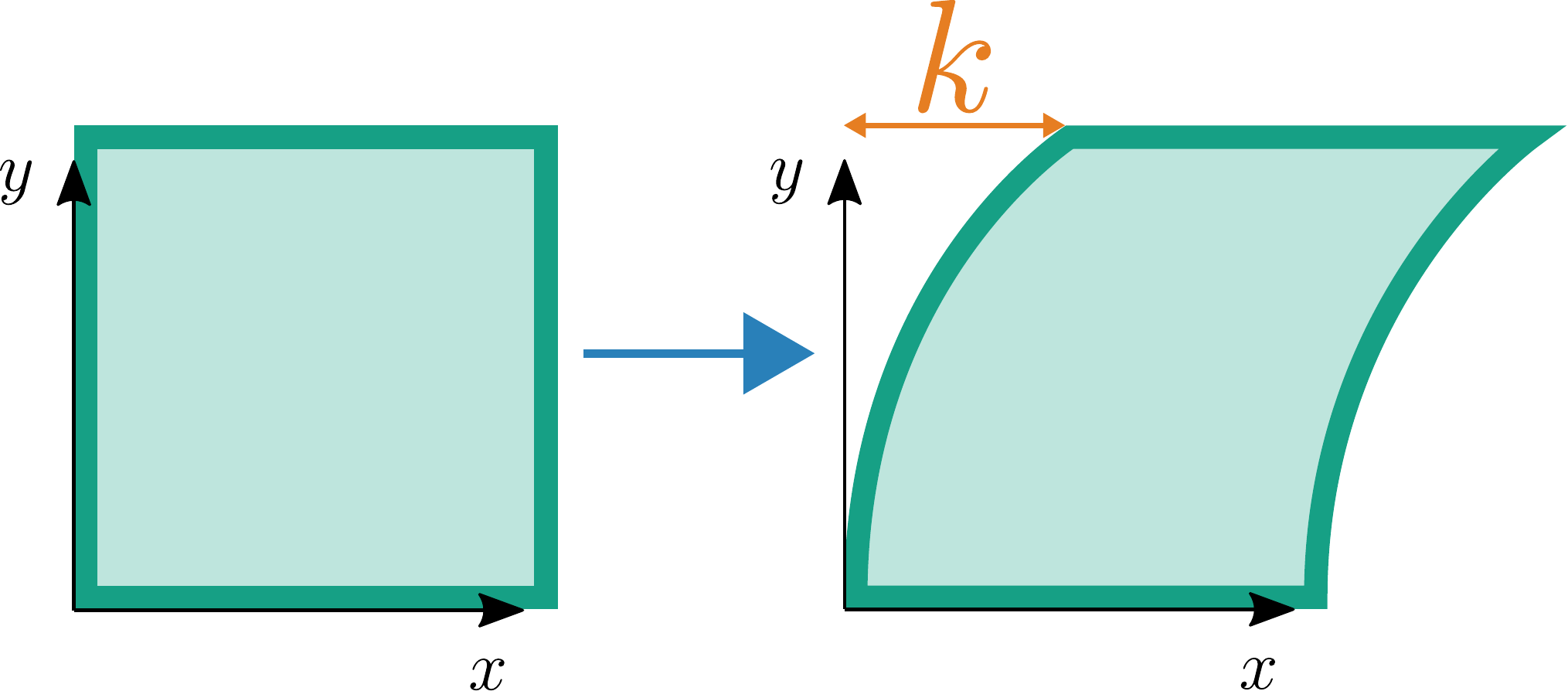}
\caption{2D plane cut of a generalized shear deformation of a unit square. An $x$-displacement of $y^{2} \cdot k$ is applied to the boundary of the cube.}
\label{fig:generalized_shear}
\end{figure}

\medbreak \noindent \textbf{Generalized shear} The generalized shear deformation test is similar to the simple shear deformation~\cite{Mihai2013}. The initial geometry is a unit cube with prescribed Dirichlet boundary conditions $\boldsymbol{u}_0 = (y^2 \cdot k, 0, 0)^T$ with $y$ the $y$-coordinate and $k$ a constant, as illustrated in figure \ref{fig:generalized_shear}.

For generalized shear deformation, the deformation gradient is equal to
\begin{equation} \label{eq:31}
\boldsymbol{F} =
\begin{pmatrix}
1 & 2ky & 0 \\
0 & 1 & 0 \\
0 & 0 & 1
\end{pmatrix}.
\end{equation}

In the same manner as in the simple shear deformation, we consider a cube made of a Mooney-Rivlin material and can apply the same methods to find the analytical strain energy density function $\boldsymbol{\psi}$ and the Cauchy stress tensor components $\boldsymbol{\sigma}$
\begin{align} \label{eq:32}
\begin{split}
\psi &= \int_{0}^{1} [C_{1}(\overline{I_{1}}-3)+C_{2}(\overline{II_{1}}-3)] \; \mathrm{dy} \\
& = \int_{0}^{1} 4k^{2}y^{2}(C_{1}+C_{2}) \; \mathrm{dy} \\
& = \frac{4k^{2}(C_{1}+C_{2})}{3},
\end{split}
\end{align}
\begin{align}
\begin{split}
&\sigma_{00} = \frac{k^{2}(8C_{2}+16C_{1})}{9}, \\
&\sigma_{11} = -\frac{k^{2}(16C_{2}+8C_{1})}{9}, \\
&\sigma_{22} = \frac{k^{2}(8C_{2}-8C_{1})}{9}, \\
&\sigma_{01} = k(2C_{2}+2C_{1}), \\
&\sigma_{02} = \sigma_{12} = 0. \\
\end{split}
\end{align}
We realize the same tests as the simple shear (different $k$ values, mesh precision, and mechanical parameters) and evaluate the identical quantities, $\psi$ and $\boldsymbol{\sigma}$ values. We observed an impact of the mesh on the strain energy and the Cauchy stress. The error quickly decreases on mesh refinement to reach relative errors under 2\%.

\subsection{Verification of the inverse simulation}
This section presents a series of tests to verify the consistency of our inverse method with the direct approach. More precisely, we show that the undeformed configuration corresponds to the initial configuration used to deform it. During these tests, we also compare our method to the IGA method presented in section \ref{iga} and evaluate their performance and convergence rates.

\subsubsection{Inverse shear deformation}
This test is based on the direct shear deformation verification performed in section \ref{simple_shear}. 
We verify that the inverse deformation of the simple shear and the generalized shear is consistent with the direct finite element analysis. The idea is to start the test with the deformed configuration and apply the inverse deformation to verify that the rest-configuration corresponds to the initial geometry of the direct deformation. Since both shear deformations are entirely determined by a displacement field, the inverse deformation consists of applying the opposite displacement field. It is then trivial to claim that the geometry will be recovered, i.e. a unit cube. However, this test also verifies the deformation gradient, the strain energy, and stress tensors are sound. As explained previously, those measures should be equal in both inverse and direct deformation. We verify these statements numerically in these tests.
	
\begin{figure}[H]
\includegraphics[width=\textwidth]{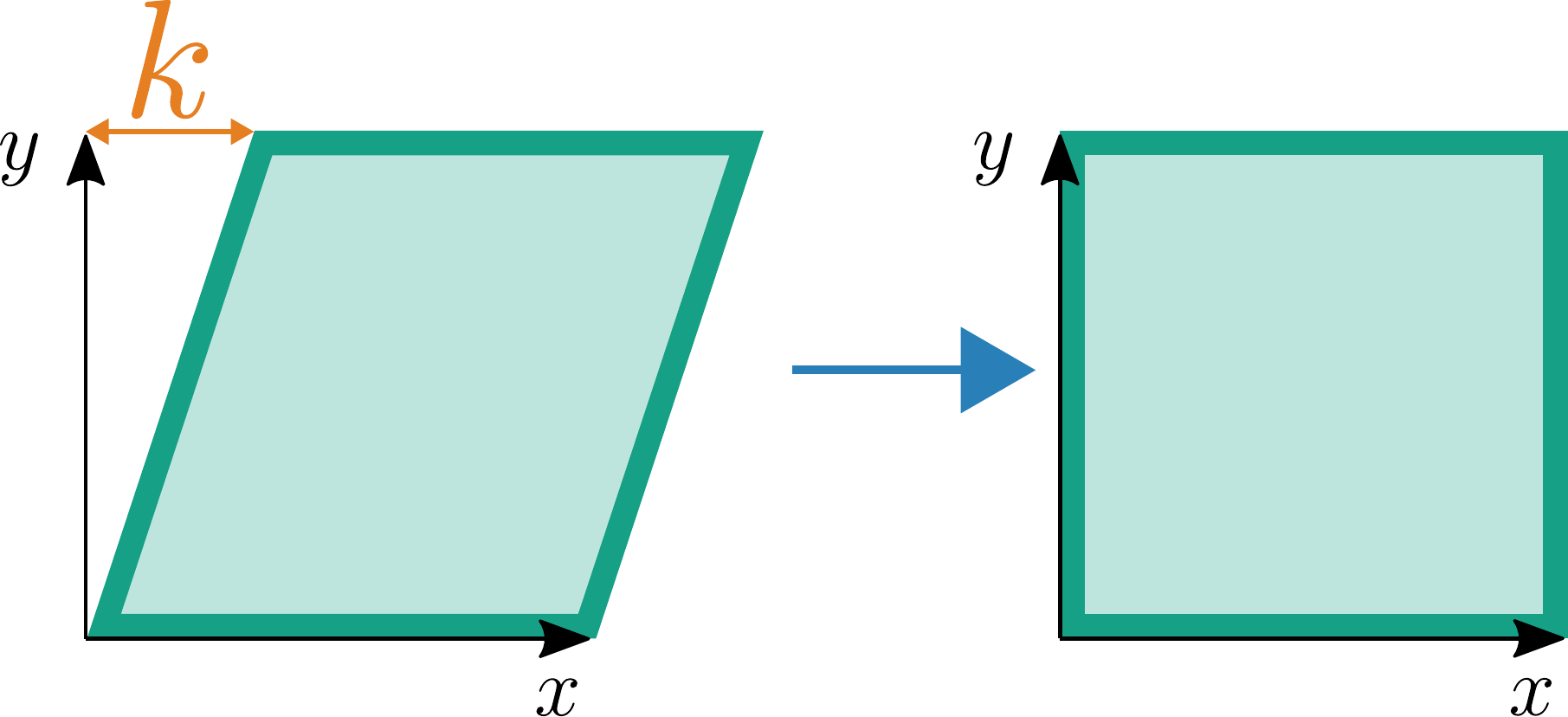}
\caption{2D plane cut of an inverse simple shear deformation of a unit cube. An $x$-displacement of $-y \cdot k$ is applied on the boundary.}
\label{fig:inverse_simple_shear}
\end{figure}

\medbreak
\noindent\textbf{Inverse simple shear} : For the inverse simple shear deformation, the material points are now shifted by $ -k \cdot y$ on the $x$-axis while the bottom is fixed ($y=0$). As illustrated in figure \ref{fig:inverse_simple_shear}.

We calculate the deformation gradient $\boldsymbol{F}$ which is equal to the deformation gradient in equation \ref{eq:28}, as expected
\begin{equation} \label{eq:37}
\nabla \boldsymbol{u'} + \boldsymbol{I} = 
\begin{pmatrix}
1 & -k & 0 \\
0 & 1 & 0 \\
0 & 0 & 1
\end{pmatrix},
\end{equation}
\begin{equation} \label{eq:38}
\boldsymbol{F} =  (\nabla \boldsymbol{u'} + \boldsymbol{I})^{-1} = 
\begin{pmatrix}
1 & k & 0 \\
0 & 1 & 0 \\
0 & 0 & 1
\end{pmatrix}.
\end{equation}
Therefore, the strain energy, which is usually defined depending on $\boldsymbol{F}$, is equal to the strain energy in equation \ref{eq:29}, and the stress tensor of equation \ref{eq:30} remains valid.
Since the deformation is homogeneous (constant deformation gradient), our quadratic finite element method is able to reproduce the analytical solution down to machine precision.

\medbreak
\noindent\textbf{Inverse generalized shear}: Similarly, the inverse version of the generalized shear deformation leads to the same deformation gradient tensor (equation \ref{eq:31}), then to the same strain energy density function (equation \ref{eq:32}). The relative error is evaluated with different discretizations of the initial mesh but the same parameters set and we obtain with high precision the initial geometry.

\subsubsection{Single tetrahedron}
\noindent \textbf{Part I}: Let us consider a mesh with a single unit tetrahedron with a linear Lagrangian finite element space. Its domain is denoted $\Omega^{T}_{0}$. The nodal coordinates are $[0 ,0, 0]^T, [1 , 0, 0]^T, [0 ,1, 0]^T$ and $[0,0, 1]^T$. The nodes with $y = 0$ are fixed, leaving only one free node. A uniform force $\boldsymbol{f}$ is applied along the $y$-axis. The tetrahedron is deformed so that the free node moves along the $y$-axis. 

\begin{figure}[H]
\includegraphics[width=\textwidth]{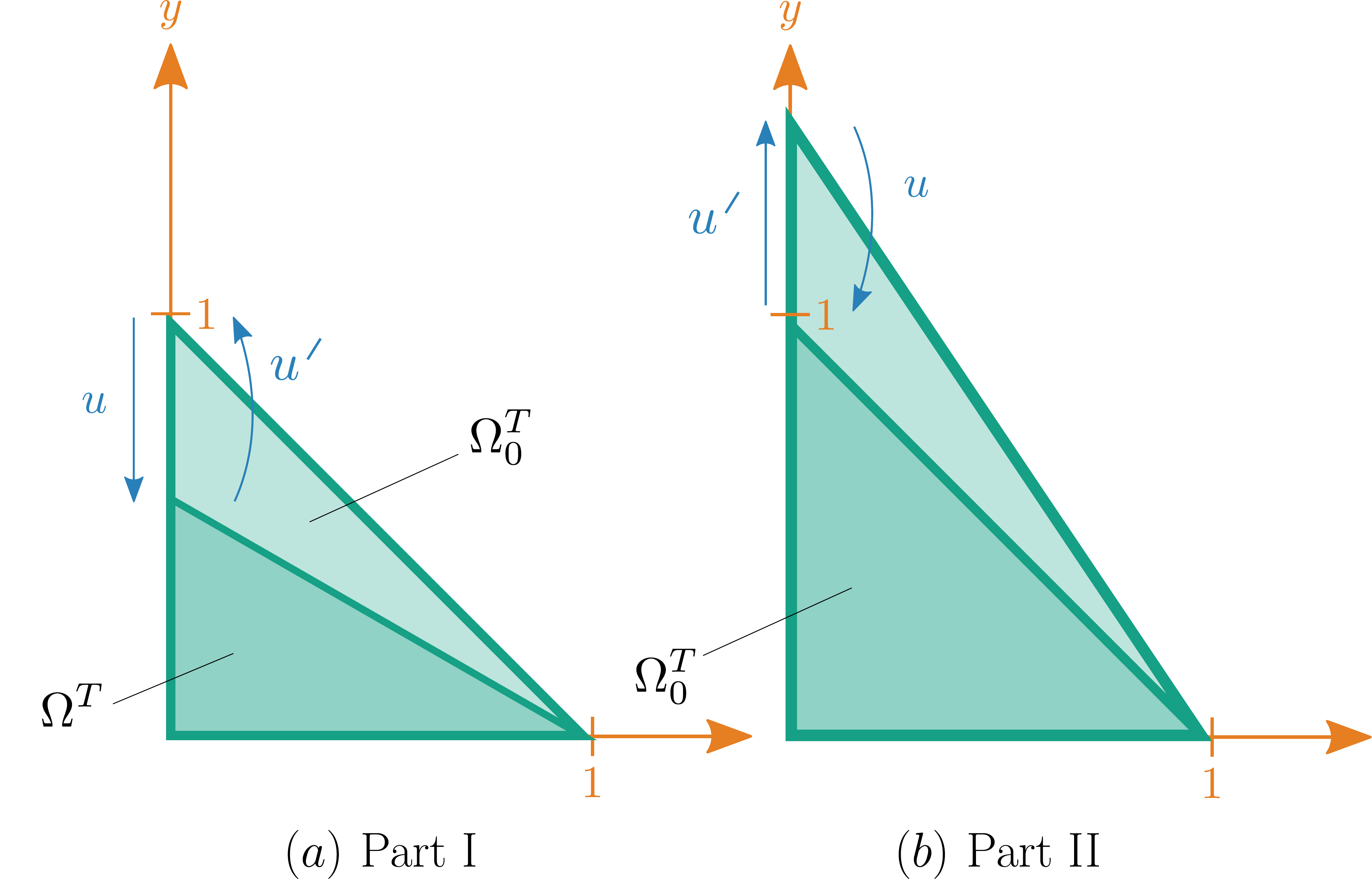}
\caption{Single tetrahedron test, plane view.}
\label{fig:tetra}
\end{figure}

In a first step, we compute the deformation $\boldsymbol{\phi}$ with the direct method. A displacement $\boldsymbol{u}$ is computed for the free node. The deformed domain is $\Omega^{T} = \boldsymbol{\phi} (\Omega^{T}_{0})$. In a second step, the initial geometry is the deformed geometry $\Omega^{T}$, i.e. a unit tetrahedron with the nodes $y = 0$ fixed, and the remaining node displaced from $\boldsymbol{u}$. The same uniform force $\boldsymbol{f'} = \boldsymbol{f}$ is applied. An inverse simulation is computed so that the displacement of the free node is $\boldsymbol{u'}$. This example is depicted in figure \ref{fig:tetra}a.

\medbreak
\noindent \textbf{Part II}: We consider the same unit tetrahedron, with the same boundary conditions. A uniform force $\boldsymbol{f'}$ is applied along the $y$-axis. 

In the first step, an inverse simulation is computed, leading to a displacement of $\boldsymbol{u'}$. In the second step, the resulting geometry is deformed with a direct simulation leading to a displacement of $\boldsymbol{u}$. This part of the example is depicted in \ref{fig:tetra}b.

The difference with the first part of the test is the order of the successive simulations. In \textbf{part I}, the inverse simulation is performed after the direct simulation. In \textbf{part II}, it is the opposite. In both parts of the test, the goal is to verify that the following relationship: $\boldsymbol{u'} = -\boldsymbol{u}$.

Furthermore, the inverse simulation is computed with IGA to compare the results and performance with our method. In this test, the error measure is defined as: $\|\boldsymbol{u'}+\boldsymbol{u}\|_{l^2}$. We measured this error with different constitutive equations and varying their associated mechanical parameters. In total, we performed $153$ tests and provided a statistical analysis in table \ref{tab:tab1}. 

\begin{table}
\centering
\begin{tabular}{|c|c|c|c|c|c|c|}
\hline
\multirow{1}{*}{} &
\multicolumn{3}{c}{Part I} &
\multicolumn{3}{c|}{Part II} \\
& PB & IGA (1) & IGA (2) & PB & IGA (1) & IGA (2) \\
\hline
average error & 4.49E-12 & 2.12E-6 & 2.28E-12 & 5.22E-12 & 1.98E-6 & 2.28E-12 \\
SD & 1.07E-11 & 1.09E-6 & 6.69E-12 & 1.25E-11 & 1.05E-6 & 2.41E-12 \\
minimum & 9.26E-22 & 5.44E-8 & 4.15E-35 & 1.04E-21 & 5.47E-8 & 6.76E-12 \\
maximum & 5.52E-11 & 4.48E-6 & 5.11E-11 & 7.26E-11 & 3.99E-6 & 5.04E-11 \\
\hline
avg \#iterations & - & 4.70 & 11.2 & - & 4.84 & 13.4 \\
avg time (ms) & 34 & 162 & 387 & 33 & 162 & 387 \\
avg time ratio & 1 & 4.75 & 11.35 & 1 & 4.70 & 11.70 \\
\hline
\end{tabular}
\caption{Benchmark results on a single tetrahedron simulation. We compare our physics-based method (PB), with the iterative geometric algorithm (IGA (1)) with the arbitrary convergence criterion $10^{-6}$, and with the IGA at the same accuracy than PB (IGA (2)).}
\label{tab:tab1}
\end{table}
We observe that the accuracy of the iterative algorithm depends on the number of iterations, but it also increases the computational cost because each iteration calls a direct simulation. Our method provides high accuracy while requiring only the solution of a problem with similar complexity to a single iteration of IGA. Beyond the numerical results, one point is that in $7$ tests over the $153$ of the \textbf{part II}, the iterative algorithm was not able to reach the accuracy of our method within $50$ iterations.

\section{Experimental results}
In this section, we will demonstrate that our inverse simulation method can match the outcome of a real experiment and therefore has value as a predictive modelling tool. 

\begin{figure}[H]
\includegraphics[width=\textwidth]{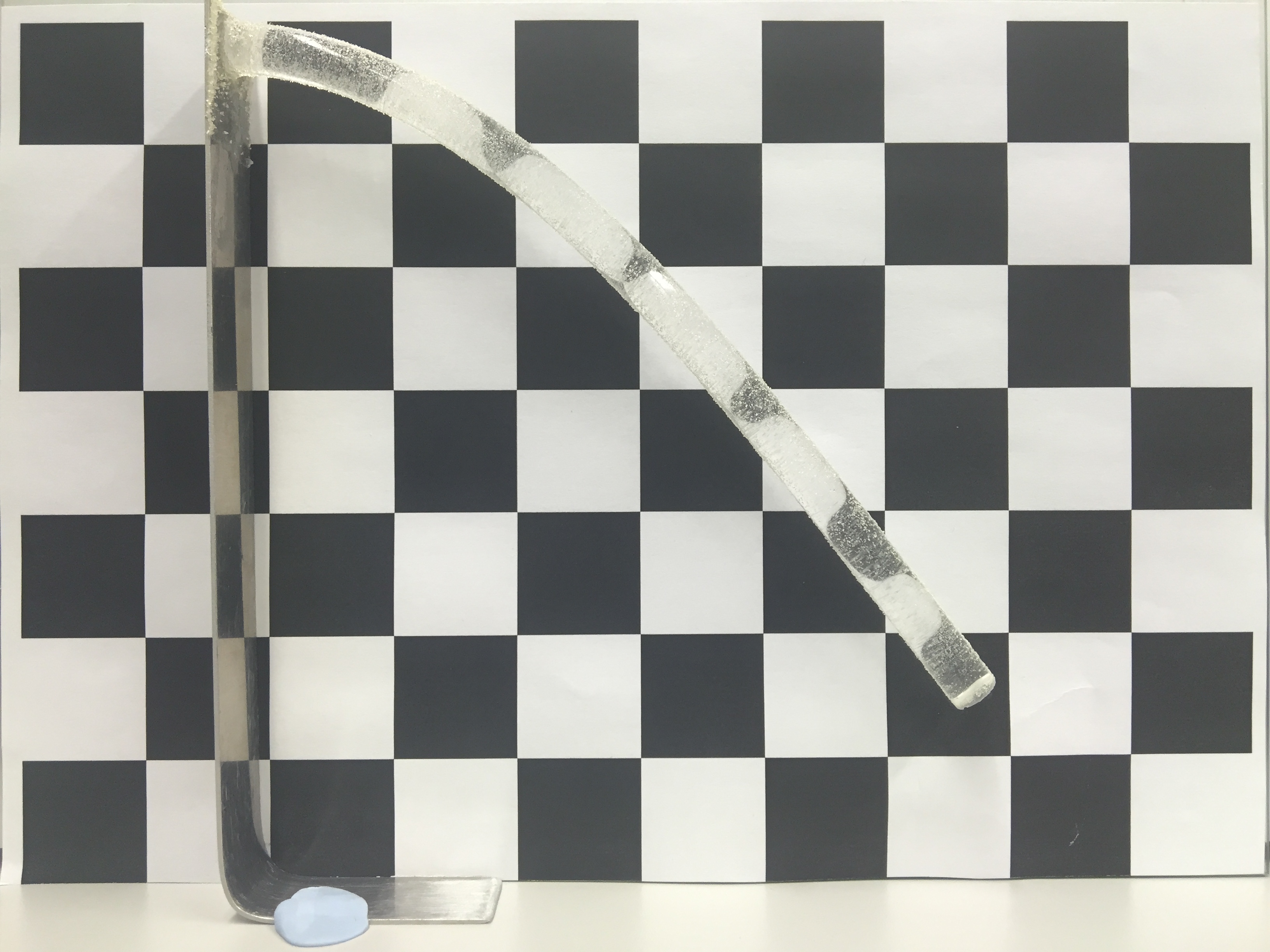}
\caption{Experimental set-up: Initially straight PDMS beam clamped on the left side and deformed by gravity.}
\label{fig:PDMS_cylinder}
\end{figure}

\begin{figure}[H]
\begin{center}
\includegraphics[width=0.5\textwidth]{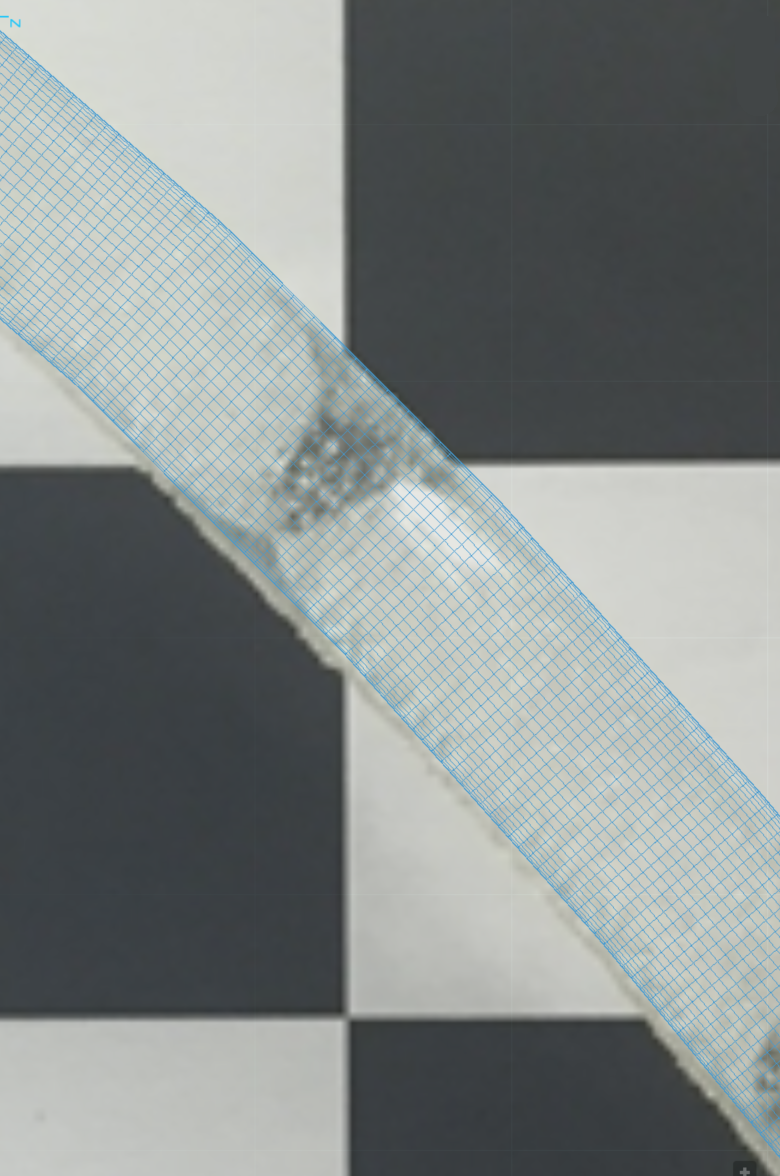}
\end{center}
\caption{Manual process in Blender to delineate the contours and extract the 3D mesh of the deformed configuration.}
\label{fig:model_from_mesh}
\end{figure}

We fixed one extremity of a beam made from Polydimethylsiloxane (PDMS) to a vertical support and allowed it to deform under gravity as shown in figure \ref{fig:PDMS_cylinder}).
To extract the mesh of the deformed configuration from the image, we used the software Blender\footnote{https://www.blender.org/} and contoured the beam on 2D images by hand, as shown in figure \ref{fig:model_from_mesh}. This mesh will be called the "reference" and used as ground-truth for this section.

To run the inverse deformation algorithm, we need three input parameters: the applied force field, the deformed configuration, and the mechanical properties. In this section, the force field is gravity and the deformed configuration was obtained by manual processing. A separate experiment was performed to obtain the mechanical properties and will be detailed in the following section.

\subsection{Material}
We used a PDMS (Sylgard 184, Ellsworth Adhesives) cylinder of density $965$ \si{kg/m^{3}} of undeformed dimensions $182$ \si{mm} and $8.5$ \si{mm} for length and diameter, respectively.

\begin{figure}[H]
\begin{center}
\includegraphics[width=0.5\textwidth]{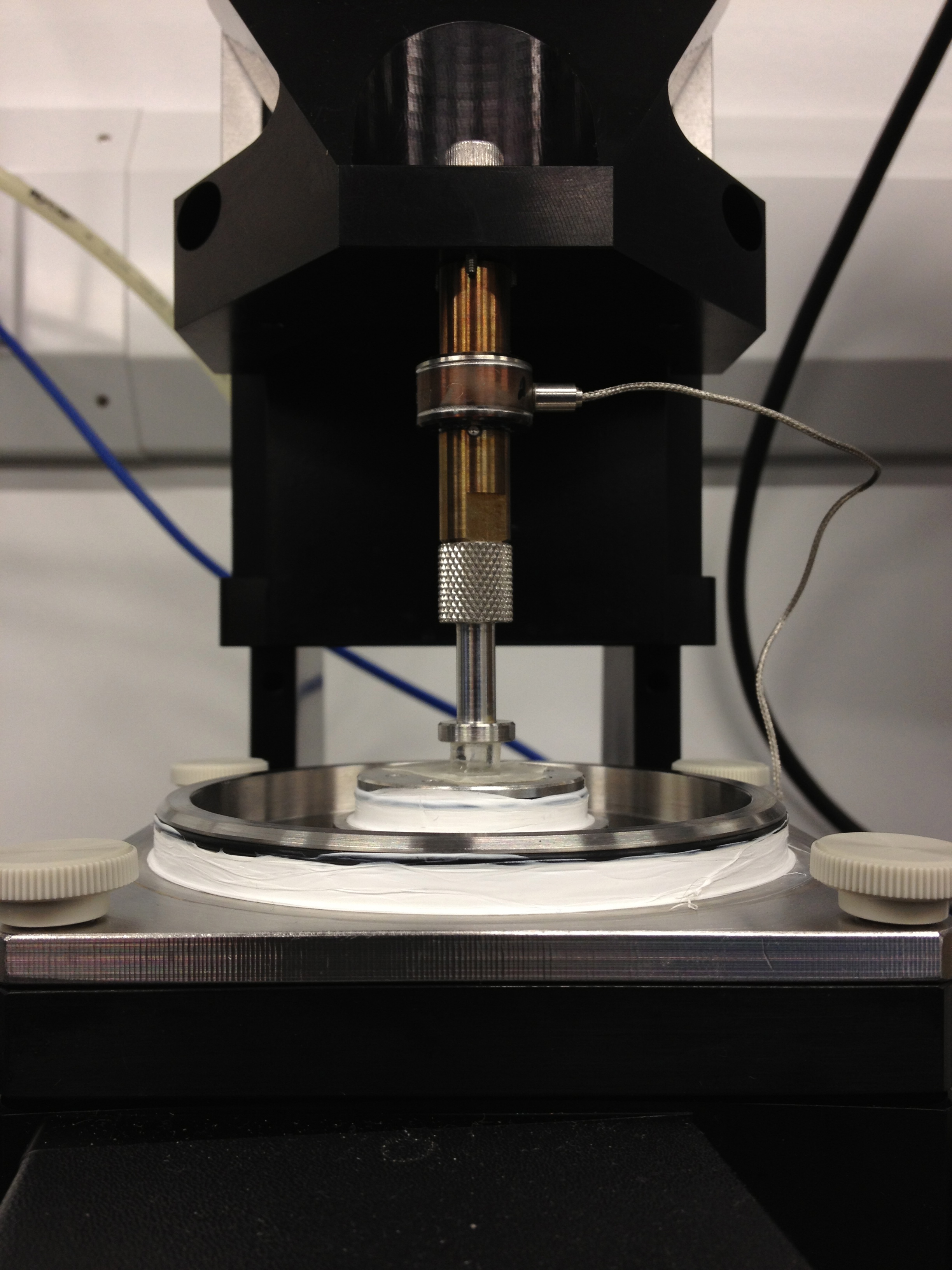}
\end{center}
\caption{Biomomentum Mach-1™ mechanical testing system used for characterizing mechanical properties of the PDMS.}
\label{fig:biomomentum}
\end{figure}

For the sample preparation the elastomeric part and curing agent were mixed in a $10:1$ ratio and cured at room temperature for $24$ \si{h} before being tested~\cite{Forte2016}.
A surgical knife was used for cutting cylindrical shapes from the second cylinder of PDMS, for compression tests samples (diameter $11$ \si{mm}, height $7 \pm 1$ \si{mm} in figure \ref{fig:biomomentum}). 

To characterize the material properties, we used the Mach-1™ mechanical testing system (Biomomentum, Canada) as a testing rig for the unconfined compression tests. We used the following protocol: 
\begin{itemize}
    \item A $1.5$ \si{mm} single-axis load cell with a resolution of $75$ \si{\micro N} was used to measure the vertical force.
    \item The vertical displacement was measured by the moving stage of the rig with a resolution of $0.1$ \si{\micro N}.
    \item To minimize friction, paraffin oil was used between the sample and the compression platens.
    \item One loading cycle was executed on each specimen. To detect the response of the material at large strains, the samples were compressed at a constant speed of $0.083$ \si{mm/s} until a displacement corresponding to 30$\%$ of the measured height was achieved. Particular attention was used to monitor the samples that had uniformly expanded in the radial direction and that their upper and lower faces remained adhered to the moving platen and the fixed platform for the entire duration of the test. 
    \item The Abaqus evaluation routine was used to fit the true stress - true strain experimental curves with a Mooney-Rivlin model. Abaqus employs a linear least-squares fit for the Mooney-Rivlin form to find the optimal model parameters.
\end{itemize}

In our case the optimal parameters are:
$ D_{1} = 7.965272689 \times 10^{-8}$ \si{\Pa}, $C_{10} = 101709.668$ \si{\Pa}, $C_{01} = 151065.460$ \si{\Pa}.

\subsection{Direct simulation}

\begin{figure}[H]
\includegraphics[width=\textwidth]{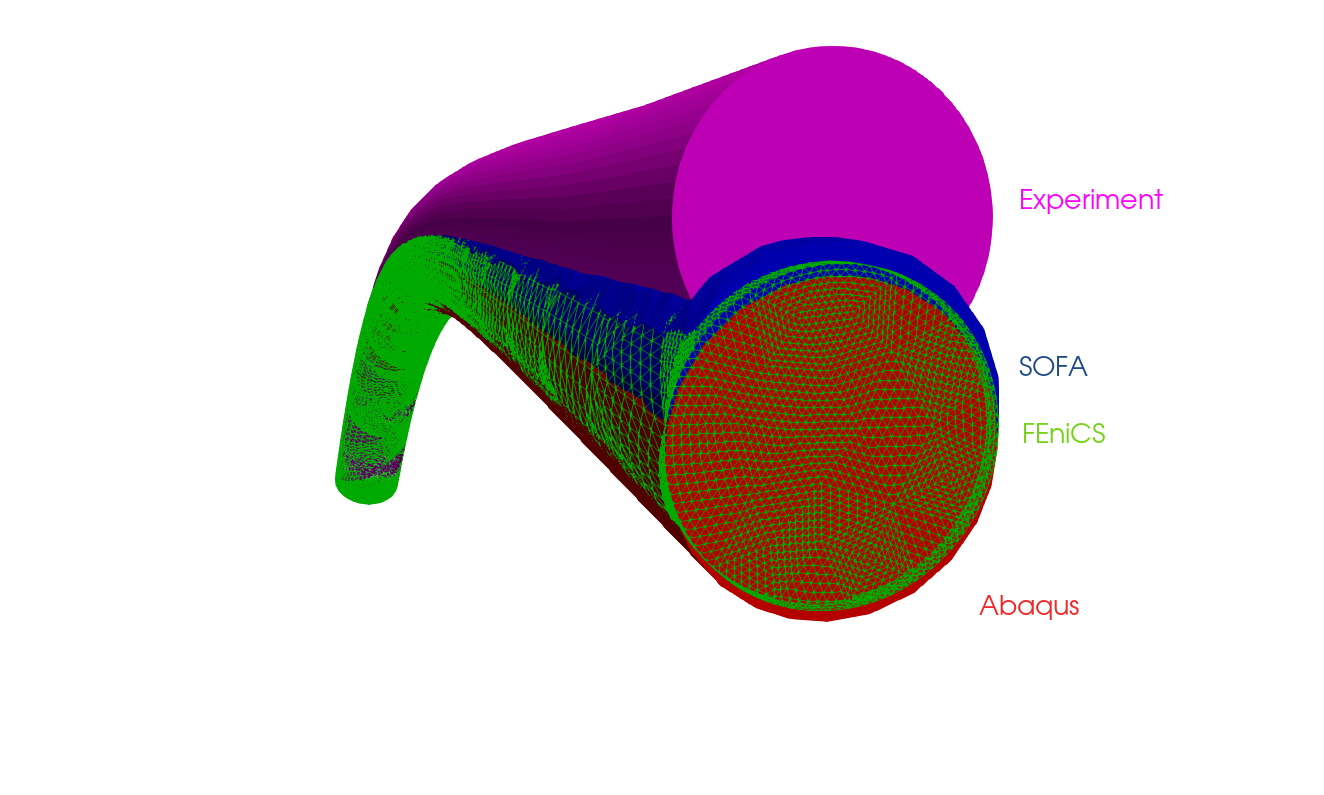}
\caption{3D plot of the forward simulation. From top to down, in magenta: the experimental data, in blue: the SOFA simulation, in wire-frame green: the FEniCS simulation and in red: the Abaqus simulation.}
\label{fig:forward-experiment}
\end{figure}

To verify the mechanical properties we compare the output of three different simulation softwares all using an incompressible Mooney-Rivlin model and boundary conditions imitating the setup shown in figure \ref{fig:forward-experiment}.

\medbreak
\noindent\textbf{FEniCS:} We used the same model as described in section \ref{fenics-model}.
\medbreak
\noindent\textbf{Abaqus:} We use a static step with a gravity load to solve the beam deformation in Abaqus. Abaqus/Standard uses Newton's method as a numerical technique for solving the nonlinear equilibrium equations. We employed  C3D8RH elements, an 8-node linear brick, hybrid/mixed, constant pressure, reduced integration with hourglass control. The hybrid/mixed formulation is needed because of the material's near-incompressibility.
\medbreak
\noindent\textbf{SOFA:} We employed the \emph{Multiplicative Jacobian Energy Decomposition} method (MJED) which is an optimized algorithm for building the stiffness and tangent stiffness matrices of non-linear hyperelastic materials~\cite{Marchesseau2010}. An MJED implementation is available in SOFA~\cite{faure2012sofa} for finite element formulation using linear tetrahedral elements. The linear system of equations was solved in every step of quasi-static simulation using a fast in-house linear equation solver based on the Cholesky decomposition.
\medbreak

\begin{figure}[H]
\includegraphics[width=\textwidth]{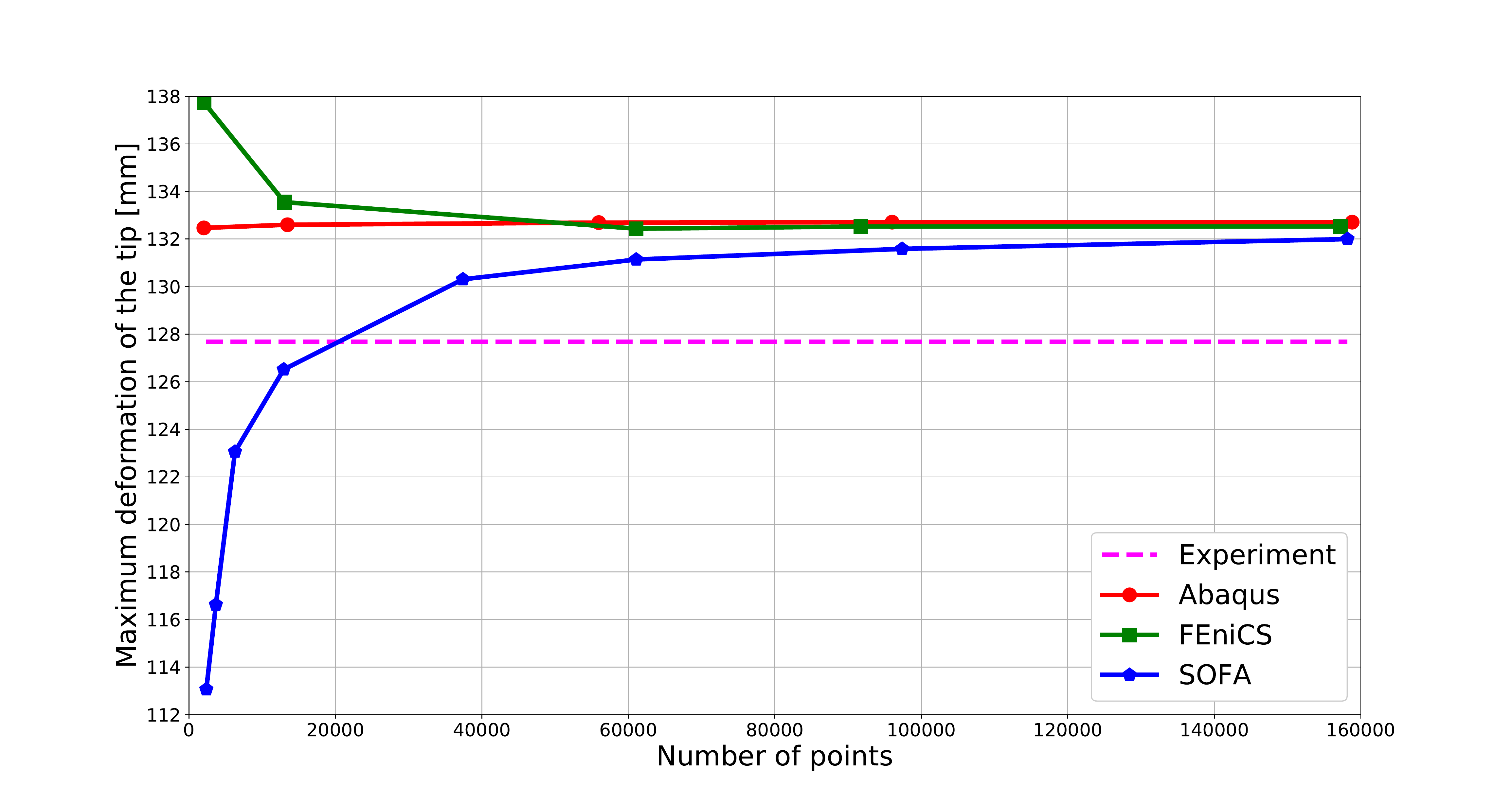}
\caption{Mesh convergence analysis of the forward simulation. We calculate the maximum deformation of the tip of beam for several level of refinement of the mesh.}
\label{fig:mesh-convergence}
\end{figure}

For each model, we perform a mesh convergence analysis shown in figure \ref{fig:mesh-convergence} where we plot the maximum deformation of the beam (located at the tip) for different mesh resolutions.

We observe that the tip displacement for the three software converge to similar solutions (FEniCS: $132.52$ \si{mm}, Abaqus: $132.71$ \si{mm}, SOFA: $130.31$ \si{mm}) while the experimental value is $127.68$ \si{mm}. We observe a small difference between the numerical solutions and the experiment.

FEniCS and Abaqus give similar results while SOFA is $2$ \si{mm} off. We observe in figure \ref{fig:mesh-convergence} that FEniCS and Abaqus converged with 60,000 points while SOFA is still not converged with \num{160000} points. One reason is that SOFA is usually designed for real-time simulation and only uses dynamic solvers which can lead to inaccuracy compared with static solvers from FEniCS and Abaqus. Furthermore, the differences between numerical solutions can be explained by the use of three slightly different formulations of the Mooney-Rivlin law as well as different solvers for solving the equation.

Some factors can explain the difference between the numerical solutions and the experimental value. For instance, the variation may be explained by inadequate constitutive equations or boundary conditions. Then, uncertainties in the mechanical properties measures may also be a factor, especially because the PDMS might exhibit slightly asymmetric behaviour under compression and tension. Finally, we obtained the reference mesh of the undeformed configuration manually based on 2D imaging data where inaccuracies can be introduced.

\subsection{Inverse simulation}
In the previous section, we compared the forward simulations of three different software with our experimental solution.
In this section, we want to verify the possibility of retrieving the undeformed configuration of our experimental solution knowing only the surface of the deformed configuration, the known applied loads and the material properties.

For this, we converted our experimental surface mesh of the deformed configuration into a volumetric mesh and applied our inverse deformation algorithm implemented using FEniCS. We previously showed a deformation difference of $4.84$ \si{mm} for the forward simulation in FEniCS. Of course, do not expect to obtain a perfectly straight beam (the ideal undeformed configuration), but rather an error on the same order as in the forward simulation.

\begin{figure}[H]
\includegraphics[width=\textwidth]{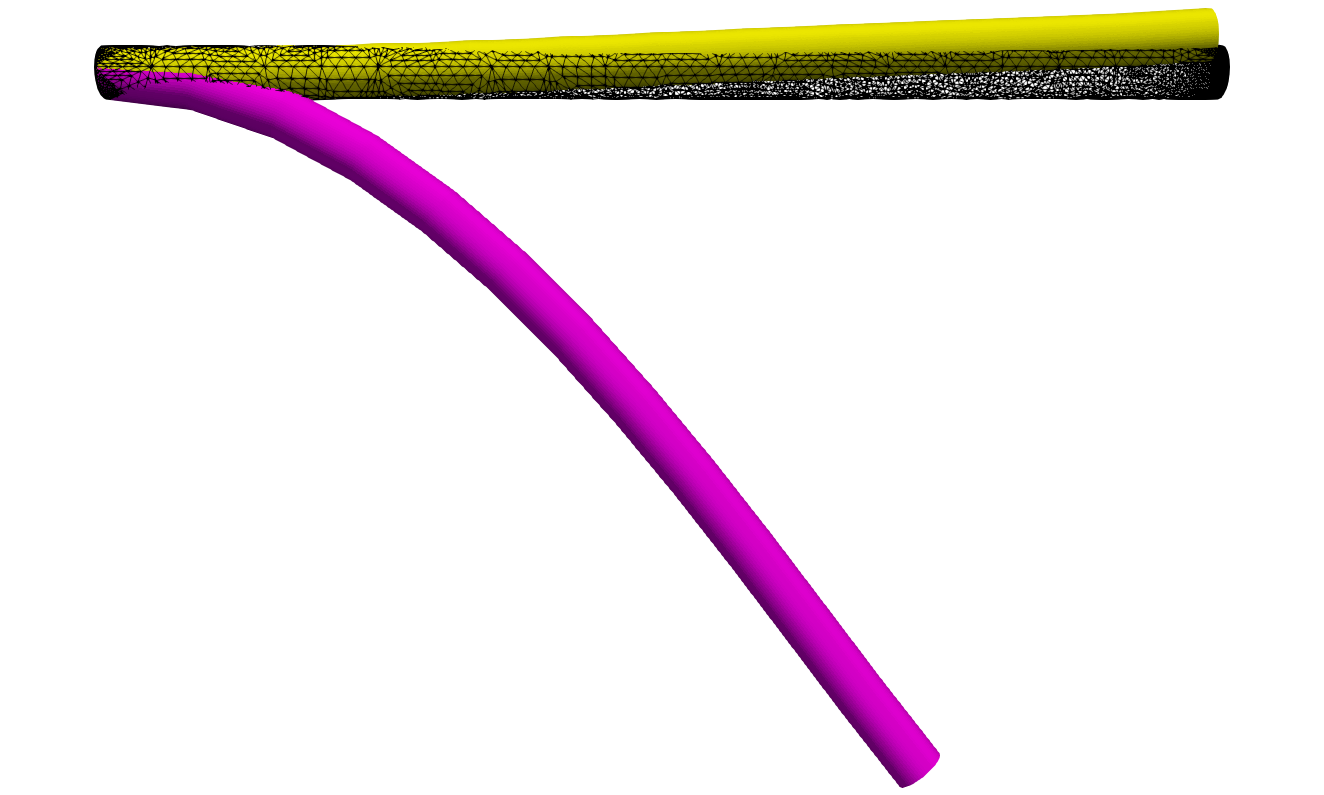}
\caption{3D plot of the inverse simulation. From down to top, in magenta: the experimental data we wish to retrieve the undeformed configuration, in wire-frame black: the theoretical straight beam , in yellow: the result of the FEniCS inverse simulation.}
\label{fig:inverse-deformation-exeriment}
\end{figure}

We show in figure \ref{fig:inverse-deformation-exeriment} the result of the inverse deformation algorithm.  As expected, the inverse simulation (in yellow) applied to the experimental data (deformed configuration in magenta) is slightly different from the theoretical straight beam that we should obtain (in black). To be more precise, we achieve an error of $5.36$ \si{mm} compared with the idealised straight beam. As mentioned previously, we expect an error on the order of that for the standard deformation problem ($4.84$ \si{mm}) due to the inherent parametric and modeling uncertainties (material model, material properties, boundary conditions, geometry) already discussed. We therefore judge that the proposed methodology has strong potential for prediction of the undeformed configuration of a soft body. 

\section{Conclusions}
In the present paper we performed a numerical and experimental study of the \emph{inverse deformation problem}.

Our study used the Lagrangian formulation of~\cite{Fachinotti2009} as a basis for implementing the inverse algorithm in the FEniCS Project finite element software. We took advantage of the automatic differentiation and code generation capabilities to bypass the difficulties of deriving and implementing the consistent Jacobian. The user must then supply the deformed configuration, the mechanical properties, boundary conditions and the applied forces. The user can easily modify the input mesh, run the code efficiently in parallel, change the constitutive model or change the boundary conditions according to their needs. We have made the code and data available in the supplementary material.

We applied the approach to simple academic examples where we considered two different incompressible hyperelastic models (neo-Hookean and Mooney-Rivlin) and different boundary conditions. We demonstrated on a simple test case that our method is more efficient in terms of robustness and accuracy than the IGA method of~\cite{Sellier2011}. We have only compared with the classical IGA method of Sellier but other works like~\cite{Rausch2017} have improved on this algorithm. However, we can say that unless an iterative approach requires only one forward model solution, in most circumstances the mechanics-based approach detailed here is likely to be faster and more robust.

Finally we applied the method to an experiment with a PDMS beam deformed under gravity. We verified and quantified the performance of the direct simulations of three different widely-used software (Abaqus, FEniCS, SOFA). Using the inverse deformation algorithm we achieve an error of $5.36$ \si{mm} for the tip displacement compared to the idealised straight beam.

Despite our progress in providing a flexible inverse deformation algorithm, some work remains to assess its robustness. Our experiments were only focused on using homogeneous nearly-incompressible hyperelastic models. Other works such as~\cite{Fachinotti2009} were interested in more complex behaviors like anisotropy. Similarly, our experiments were only based on simple geometries and more complex geometries should be considered.

We showed the validity of our approach for the beam problem by generating a mesh of the deformed configuration from 2D images and recovering the undeformed configuration. In future work we intend to apply this algorithm to segmented 3D geometries to calculate the undeformed configuration of an organ. 

\section*{Supplementary material}
The reference~\cite{my_code-web-page} (\href{https://doi.org/10.6084/m9.figshare.14035793}{doi:10.6084/m9.figshare.14035793}) contains a full implementations of the forward and inverse deformation problems using the FEniCS Project finite element software. The latest version is also available on GitHub at \href{https://github.com/Ziemnono/fenics-inverseFEM}{https://github.com/Ziemnono/fenics-inverseFEM}


\begin{thebibliography}{10}
\expandafter\ifx\csname url\endcsname\relax
  \def\url#1{\texttt{#1}}\fi
\expandafter\ifx\csname urlprefix\endcsname\relax\def\urlprefix{URL }\fi
\expandafter\ifx\csname href\endcsname\relax
  \def\href#1#2{#2} \def\path#1{#1}\fi

\bibitem{Raghavan2006}
M.~L. Raghavan, B.~Ma, M.~F. Fillinger, {Non-invasive determination of
  zero-pressure geometry of arterial aneurysms}, Annals of Biomedical
  Engineering 34 (2006) 1414--1419.
\newblock \href {https://doi.org/10.1007/s10439-006-9115-7}
  {\path{doi:10.1007/s10439-006-9115-7}}.

\bibitem{Lu2007}
J.~Lu, X.~Zhou, M.~L. Raghavan, {Inverse elastostatic stress analysis in
  pre-deformed biological structures : Demonstration using abdominal aortic
  aneurysms}, Journal of Biomechanics 40 (2007) 693--696.
\newblock \href {https://doi.org/10.1016/j.jbiomech.2006.01.015}
  {\path{doi:10.1016/j.jbiomech.2006.01.015}}.

\bibitem{Mira2018}
A.~M{\^{i}}ra, A.~K. Carton, S.~Muller, Y.~Payan, {A biomechanical breast model
  evaluated with respect to MRI data collected in three different positions},
  Clinical Biomechanics 60 (2018) 191--199.
\newblock \href {http://arxiv.org/abs/1811.10221} {\path{arXiv:1811.10221}},
  \href {https://doi.org/10.1016/j.clinbiomech.2018.10.020}
  {\path{doi:10.1016/j.clinbiomech.2018.10.020}}.

\bibitem{Koishi2001}
M.~Koishi, S.~Govindjee, {Inverse design methodology of a tire}, Tire Science
  and Technology 29 (2001) 155--170.
\newblock \href {https://doi.org/10.2346/1.2135236}
  {\path{doi:10.2346/1.2135236}}.

\bibitem{Cardona2008}
V.~D. Fachinotti, A.~Cardona, P.~Jetteur, {Finite element modelling of inverse
  design problems in large deformations anisotropic hyperelasticity},
  International Journal For Numerical Methods In Engineering 74 (2008)
  894--910.
\newblock \href {https://doi.org/10.1002/nme.2193}
  {\path{doi:10.1002/nme.2193}}.

\bibitem{Zl1957}
J.~Adkins, {A reciprocal plane property of the finite plan strain equations},
  Journal of the Mechanics Physics of Solid 6 (1958) 267--275.
\newblock \href {https://doi.org/10.1016/0022-5096(58)90002-4}
  {\path{doi:10.1016/0022-5096(58)90002-4}}.

\bibitem{Schield1967}
R.~T. Schield, {Inverse deformation results in finite elasticity}, Zeitschrift
  f{\"{u}}r angewandte Mathematik und Physik ZAMP 18 (1967) 490--500.
\newblock \href {https://doi.org/10.1007/BF01601719}
  {\path{doi:10.1007/BF01601719}}.

\bibitem{Carlson1969}
D.~E. Carlson, T.~Shield, {Inverse deformation results for elastic materials},
  Zeitschrift f{\"{u}}r angewandte Mathematik und Physik ZAMP 20 (1969)
  261--263.
\newblock \href {https://doi.org/10.1007/BF01595564}
  {\path{doi:10.1007/BF01595564}}.

\bibitem{Carroll2005}
M.~M. Carroll, F.~J. Rooney, {Implications of Shield ' s inverse deformation
  theorem for compressible finite elasticity}, Zeitschrift f{\"{u}}r angewandte
  Mathematik und Physik ZAMP 56 (2005) 1048--1060.
\newblock \href {https://doi.org/10.1007/s00033-005-2023-0}
  {\path{doi:10.1007/s00033-005-2023-0}}.

\bibitem{Govindjee1996}
S.~Govindjee, P.~A. Mihalic, Computational methods for inverse finite
  elastostatics, Computer Methods in Applied Mechanics and Engineering 136
  (1996) 47--57.
\newblock \href {https://doi.org/10.1016/0045-7825(96)01045-6}
  {\path{doi:10.1016/0045-7825(96)01045-6}}.

\bibitem{Govindjee1998}
S.~Govindjee, P.~A. Mihalic, {Computational methods for inverse deformations in
  quasi-incompressible finite elasticity}, International Journal For Numerical
  Methods In Engineering 43 (1998) 821--838.
\newblock \href
  {https://doi.org/10.1002/(SICI)1097-0207(19981115)43:5<821::AID-NME453>3.0.CO;2-C}
  {\path{doi:10.1002/(SICI)1097-0207(19981115)43:5<821::AID-NME453>3.0.CO;2-C}}.

\bibitem{Yamada1998}
T.~Yamada, {Finite element procedure of initial shape determination for
  hyperelasticity}, Structural Engineering and Mechanics 6 (1998) 173--183.
\newblock \href {https://doi.org/10.12989/sem.1998.6.2.173}
  {\path{doi:10.12989/sem.1998.6.2.173}}.

\bibitem{Fachinotti2009}
A.~Albanesi, V.~Fachinotti, A.~Cardona, Design of compliant mechanisms that
  exactly fit a desired shape, Mecánica Computacional 28 (2009) 3191--3205.

\bibitem{Sellier2011}
M.~Sellier, {An iterative method for the inverse elasto-static problem},
  Journal of Fluids and Structures 27 (2011) 1461--1470.
\newblock \href {https://doi.org/10.1016/j.jfluidstructs.2011.08.002}
  {\path{doi:10.1016/j.jfluidstructs.2011.08.002}}.

\bibitem{Bols2013}
J.~Bols, J.~Degroote, B.~Trachet, B.~Verhegghe, P.~Segers, J.~Vierendeels, {A
  computational method to assess the in vivo stresses and unloaded
  configuration of patient-specific blood vessels}, Journal of Computational
  and Applied Mathematics 246 (2013) 10--17.
\newblock \href {https://doi.org/10.1016/j.cam.2012.10.034}
  {\path{doi:10.1016/j.cam.2012.10.034}}.

\bibitem{Chen2014}
X.~Chen, C.~Zheng, W.~Xu, K.~Zhou, {An asymptotic numerical method for inverse
  elastic shape design}, ACM Transactions on Graphics 33 (2014).
\newblock \href {https://doi.org/10.1145/2601097.2601189}
  {\path{doi:10.1145/2601097.2601189}}.

\bibitem{Ly2018}
M.~Ly, R.~Casati, F.~Bertails-Descoubes, M.~Skouras, L.~Boissieux, {Inverse
  elastic shell design with contact and friction}, SIGGRAPH Asia 2018 Technical
  Papers, SIGGRAPH Asia 2018 37 (2018).
\newblock \href {https://doi.org/10.1145/3272127.3275036}
  {\path{doi:10.1145/3272127.3275036}}.

\bibitem{alnaes_fenics_2015}
M.~Alnæs, J.~Blechta, J.~Hake, A.~Johansson, B.~Kehlet, A.~Logg,
  C.~Richardson, J.~Ring, M.~E. Rognes, G.~N. Wells, The {FEniCS} project
  version 1.5, Archive of Numerical Software 3 (2015).
\newblock \href {https://doi.org/10.11588/ans.2015.100.20553}
  {\path{doi:10.11588/ans.2015.100.20553}}.

\bibitem{Mihai2013}
L.~A. Mihai, A.~Goriely, {Numerical simulation of shear and the Poynting
  effects by the finite element method: An application of the generalised
  empirical inequalities in non-linear elasticity}, International Journal of
  Non-Linear Mechanics 49 (2013) 1--14.
\newblock \href {https://doi.org/10.1016/j.ijnonlinmec.2012.09.001}
  {\path{doi:10.1016/j.ijnonlinmec.2012.09.001}}.

\bibitem{Lee2017}
C.~K. Lee, L.~{Angela Mihai}, J.~S. Hale, P.~Kerfriden, S.~P. Bordas, {Strain
  smoothing for compressible and nearly-incompressible finite elasticity},
  Computers and Structures 182 (2017) 540--555.
\newblock \href {https://doi.org/10.1016/j.compstruc.2016.05.004}
  {\path{doi:10.1016/j.compstruc.2016.05.004}}.

\bibitem{alnaes_unified_2014}
M.~S. Alnæs, A.~Logg, K.~B. Ølgaard, M.~E. Rognes, G.~N. Wells, Unified form
  language: A domain-specific language for weak formulations of partial
  differential equations, {ACM} Trans. Math. Softw. 40 (2014) 9:1--9:37.
\newblock \href {https://doi.org/10.1145/2566630} {\path{doi:10.1145/2566630}}.

\bibitem{logg_ffc_2012}
A.~Logg, K.~B. Ølgaard, M.~E. Rognes, G.~N. Wells, {FFC}: the {FEniCS} form
  compiler, in: A.~Logg, K.-A. Mardal, G.~Wells (Eds.), Automated Solution of
  Differential Equations by the Finite Element Method, Lecture Notes in
  Computational Science and Engineering, Springer Berlin Heidelberg, 2012, pp.
  227--238.

\bibitem{logg_dolfin:_2010}
A.~Logg, G.~N. Wells, {DOLFIN}: Automated finite element computing, {ACM}
  Trans. Math. Softw. 37 (2010) 20:1--20:28.
\newblock \href {https://doi.org/10.1145/1731022.1731030}
  {\path{doi:10.1145/1731022.1731030}}.

\bibitem{petsc-web-page}
S.~Balay, S.~Abhyankar, M.~F. Adams, J.~Brown, P.~Brune, K.~Buschelman,
  L.~Dalcin, A.~Dener, V.~Eijkhout, W.~D. Gropp, D.~Karpeyev, D.~Kaushik, M.~G.
  Knepley, D.~A. May, L.~C. McInnes, R.~T. Mills, T.~Munson, K.~Rupp, P.~Sanan,
  B.~F. Smith, S.~Zampini, H.~Zhang, H.~Zhang, {PETS}c {W}eb page,
  \url{https://www.mcs.anl.gov/petsc} (2019).

\bibitem{my_code-web-page}
A.~Mazier, A.~Bilger, A.~E. Forte, I.~Peterlik, J.~S. Hale, S.~P. Bordas,
  Supplementary material for inverse deformation analysis: an experimental and
  numerical assessment using the {FEniCS} project,
  \url{10.6084/m9.figshare.14035793} (2021).

\bibitem{Forte2016}
A.~E. Forte, S.~Galvan, F.~Manieri, F.~{Rodriguez y Baena}, D.~Dini, {A
  composite hydrogel for brain tissue phantoms}, Materials and Design 112
  (2016) 227--238.
\newblock \href {https://doi.org/10.1016/j.matdes.2016.09.063}
  {\path{doi:10.1016/j.matdes.2016.09.063}}.

\bibitem{Marchesseau2010}
S.~Marchesseau, T.~Heimann, S.~Chatelin, R.~Willinger, H.~Delingette, Fast
  porous visco-hyperelastic soft tissue model for surgery simulation:
  Application to liver surgery, Progress in Biophysics and Molecular Biology
  103 (2010) 185--196, special Issue on Biomechanical Modelling of Soft Tissue
  Motion.
\newblock \href {https://doi.org/10.1016/j.pbiomolbio.2010.09.005}
  {\path{doi:10.1016/j.pbiomolbio.2010.09.005}}.

\bibitem{faure2012sofa}
F.~Faure, C.~Duriez, H.~Delingette, J.~Allard, B.~Gilles, S.~Marchesseau,
  H.~Talbot, H.~Courtecuisse, G.~Bousquet, I.~Peterlik, et~al., Sofa: A
  multi-model framework for interactive physical simulation, in: Soft tissue
  biomechanical modeling for computer assisted surgery, Springer, 2012, pp.
  283--321.
\newblock \href {https://doi.org/10.1007/8415\_2012\_125}
  {\path{doi:10.1007/8415\_2012\_125}}.

\bibitem{Rausch2017}
M.~K. Rausch, M.~Genet, J.~D. Humphrey, {An augmented iterative method for
  identifying a stress-free reference configuration in image-based
  biomechanical modeling}, Journal of Biomechanics 58 (2017) 227--231.
\newblock \href {https://doi.org/10.1016/j.jbiomech.2017.04.021}
  {\path{doi:10.1016/j.jbiomech.2017.04.021}}.

\end{thebibliography}
\end{document}